\documentclass[rmp,aps,twocolumn,nofootinbib,longbibliography]{revtex4-2}
\usepackage{amsmath}
\usepackage{amssymb}
\usepackage{epsfig}
\usepackage{graphicx}
\usepackage{hyperref}
\usepackage[dvipsnames]{xcolor}

\begin{document}
\title{{\it  Colloquium}: Semi-Dirac Fermions in Quantum Matter}

\author{Bruno Uchoa}
\email{uchoa@ou.edu}
\affiliation{Department of Physics and Astronomy, 
University of Oklahoma, 
Norman, Oklahoma 73019, USA}
\author{Mohamed M. Elsayed}
\affiliation{Department of Physics, 
University of Vermont, Burlington, Vermont 05405, USA}
\author{Valeri N. Kotov}
\affiliation{Department of Physics, 
University of Vermont, Burlington, Vermont 05405, USA}
\author{Yinming Shao}
\email{ymshao@psu.edu}
\affiliation{Department of Physics and Materials Research Institute, 
Pennsylvania State University, University Park, Pennsylvania 16802, USA}
\author{Dmitri N. Basov}
\affiliation{Department of Physics, 
Columbia University, New York, New York 10027, USA}

\begin{abstract}  
This article reviews the recent progress on the subject of semi-Dirac fermions, two dimensional quasiparticles that disperse quadratically, as Galilean invariant particles, or linearly, as massless relativistic particles, depending on their direction of motion. These quasiparticles exist at phase boundaries set by the continuous change in the connectivity of Fermi surfaces, known as Lifshitz transitions. The basic properties and the current experimental evidence of the existence of these quasiparticles in both synthetic lattices and quantum materials are presented. The many-body problem of semi-Dirac fermions is discussed from a theoretical perspective with an eye to physical observables of relevance to experiments.  
\end{abstract}                                                                 
 
\date{\today}
\maketitle
\tableofcontents

\section{Introduction}
\label{sec:intro} 

\begin{figure*}[!ht] 
    \centering
    \includegraphics[width=0.9\textwidth]{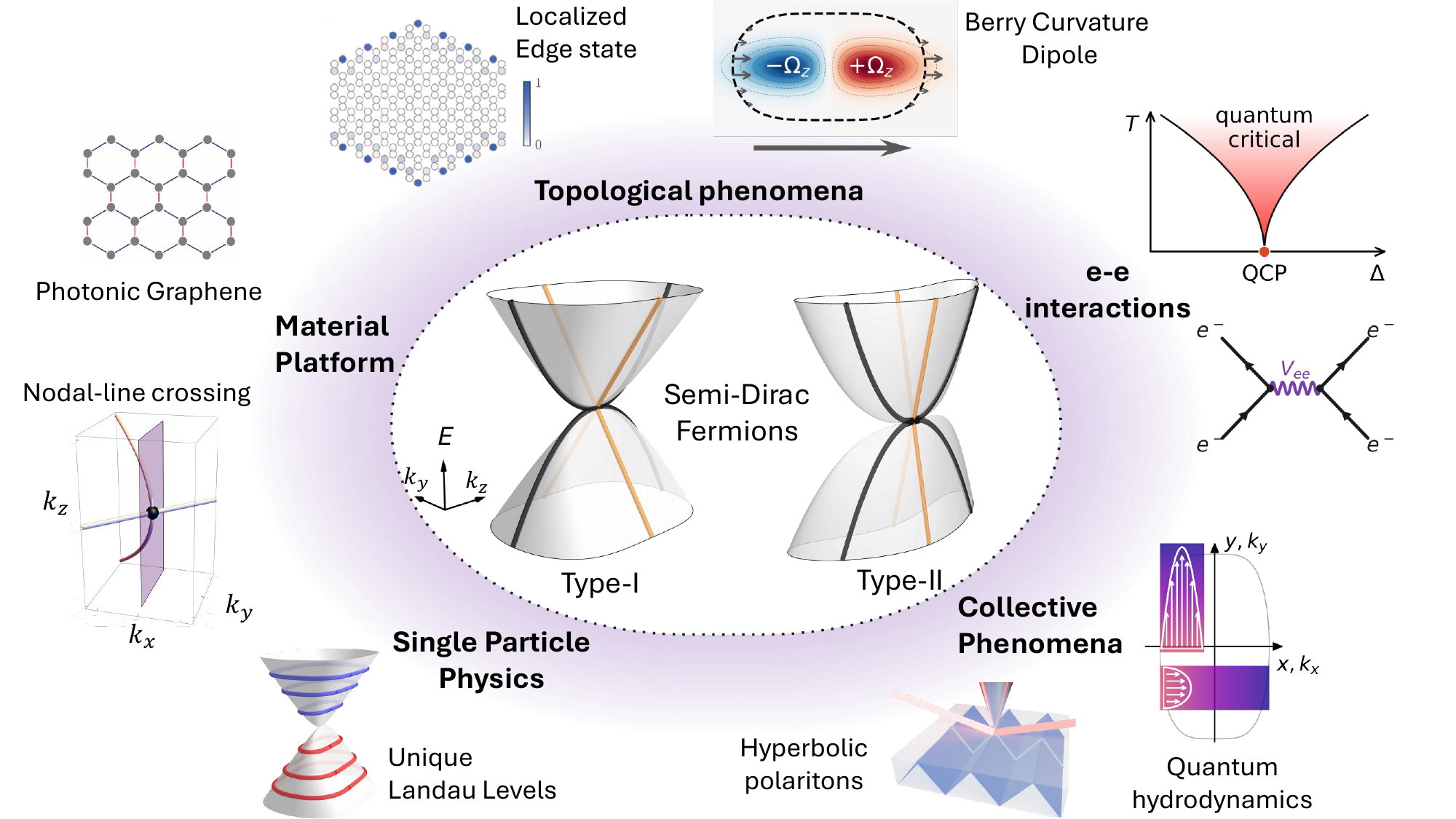}
    \caption{Overview of the topics covered in the review. Semi-Dirac particles have been experimentally observed in different platforms, including synthetic lattices of various kinds (Sec. \ref{sec:synthetic}) and in quantum materials with nodal lines (Sec. \ref{sec:solids}). These quasiparticles show distinctive optical (Sec. \ref{sec:Optical}) and 
    topological properties (Sec. \ref{sec:formation} and \ref{sec:Hall}) and have a singular spectrum of Landau levels (Sec. \ref{sec:LL}). The many-body problem resulting from electron-electron interactions is predicted to show a range of effects in both perturbative (Sec. \ref{sec:pert}) and non-perturbative regimes (Sec. \ref{sec:largeN}), such as quantum hydrodynamics (Sec. \ref{sec:hydrodynamics}), strong renormalizations (Sec. \ref{sec:manybody}) and anisotropic quantum critical phenomena (Sec. \ref{sec:instability}). In Sec. \ref{sec:Conclusion} we discuss promising directions for the field, including new many-body states, light-matter coupling in hyperbolic media and quantum geometric effects. }
    \label{fig:Overview}
\end{figure*}

Since the early days of the theory of solids \cite{Hoddeson1987}, 
it has been established that non-relativistic nearly-free electrons are Galilean invariant particles that propagate with the kinetic energy $\varepsilon(\mathbf{p}) = \mathbf{p}^2/2m$, where $\mathbf{p}$ is their momentum and $m$  their effective mass. More recently, it has been discovered theoretically \cite{DiVicenzo1984, Wallace1947}  and experimentally \cite{Novoselov2004} that the low energy quasiparticles in two dimensional (2D) crystals such as graphene behave as massless relativistic neutrinos with kinetic energy $\varepsilon(\mathbf{p})= \pm v |\mathbf{p}|$, where $\pm$ describe the particle-hole branches and $v$ is the quasiparticle velocity, analogous to the speed of light. A new development was the prediction and experimental observation of semi-Dirac fermions, 2D quasiparticles that propagate as relativistic or as Galilean invariant particles contingent on their direction of motion.  In their simplest form, semi-Dirac fermions have a hybrid kinetic energy that is neither purely Galilean invariant nor entirely relativistic,
\begin{equation}
\varepsilon_{\pm}(\mathbf{p}) = \pm \sqrt{ p_x^4/(2m)^2 +v^2 p_y^2}.
\label{SemiDirac_spectrum}
\end{equation}
The quasiparticles thus have parabolic dispersion along one direction, with effective mass $m$, and linear dispersion in a perpendicular direction with quasiparticle velocity $v$. 

Semi-Dirac fermions belong to a broader class of anisotropic quasiparticles in two and three dimensions that form in states that mediate topological transitions between different phases \cite{Yang2013}. 
The presence of semi-Dirac fermions in 2D quantum systems is a manifestation of Lifshitz transitions \cite{Lifshitz1960,Volovik2017}, which describe a continuous change in the connectivity of Fermi surfaces,  nodal points or nodal lines   through control of an external tuning parameter, such as chemical potential, pressure, etc.  
The fundamental anisotropy in the kinetic energy of these quasiparticles has important implications for various physical observables and generates a surprisingly rich landscape of phenomena in 2D involving topological phase transitions and interaction effects. One could ask: what are the distinctive optical properties for systems with semi-Dirac fermions? What is the current evidence for the existence of such particles? Do electron-electron interactions have unique unconventional effects compared to the relativistic case? What is the quantum hydrodynamic behavior of semi-Dirac fermions and their conservation laws? How does this essential anisotropy in the kinetic energy affect quantum critical phenomena? 
This Colloquium will touch upon those topics, among other ones summarized in Fig. \ref{fig:Overview}. Its purpose is not to be exhaustive but to highlight the remarkable physics that emerges from these quasiparticles in different quantum systems. 

In the following, we review the current status of this field from both the theoretical and experimental perspectives and suggest new exciting directions. 
\section{Basic properties}
\label{sec:free}

\subsection{Formation of semi-Dirac fermions}
\label{sec:formation}

Semi-Dirac fermions form at a Lifshitz transition that merges two or more Dirac points \cite{MontambauxMerging, montambaux2009universal, Montambaux2018}. The touching points between the conduction and valence bands, called semi-Dirac points, occur at a topological phase boundary between a semimetallic phase with Dirac quasiparticles and another phase that has either a trivial gap or a reduced number of Dirac points. 

\begin{figure}
            \centering
        
        \includegraphics[width=1\columnwidth]{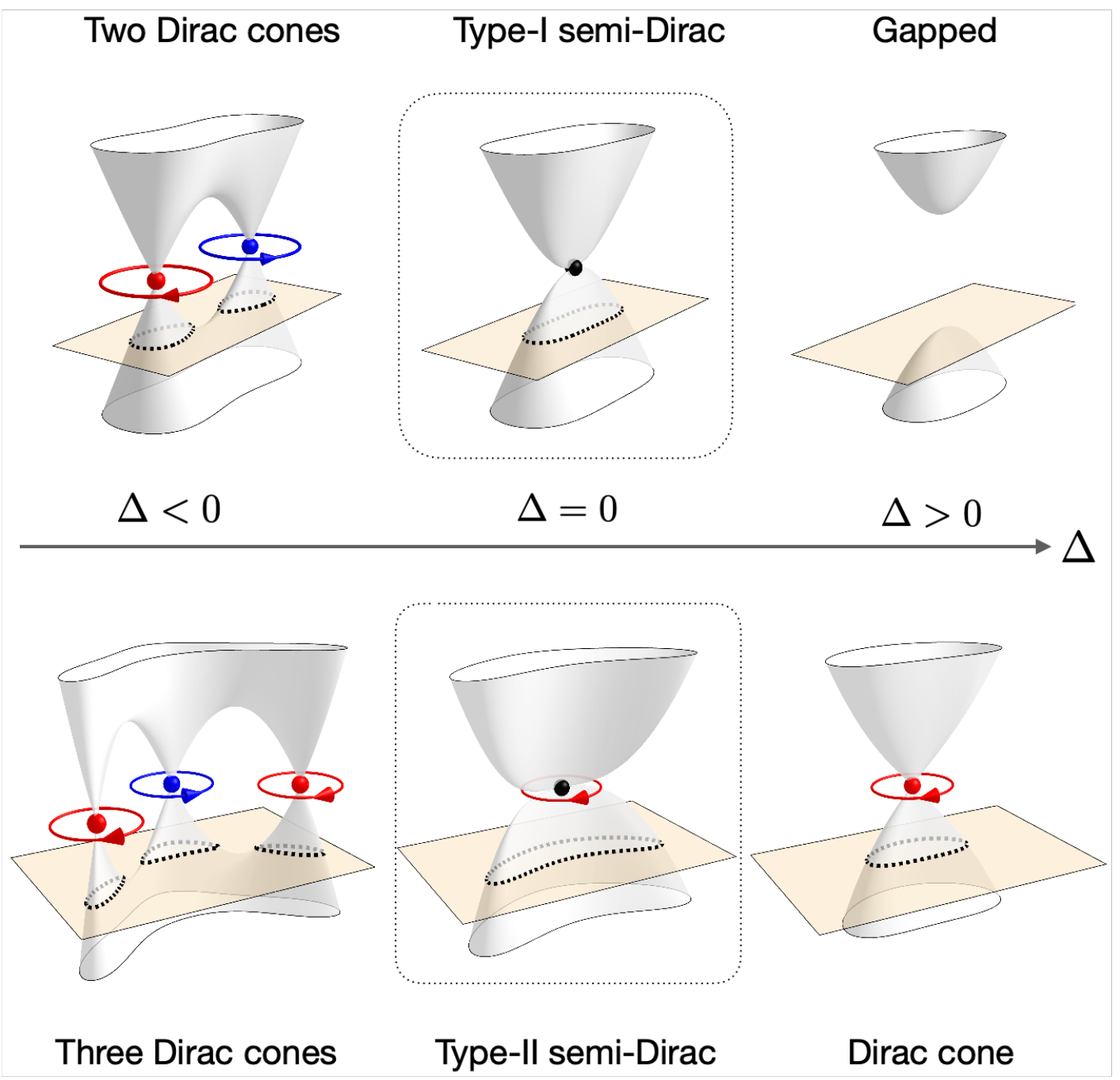}
\caption{Topological Lifshitz transition controlled by parameter $\Delta$. Top row: in the Dirac phase $\Delta<0$, the merger of two Dirac cones  with opposite $\pi$ Berry phases, indicated by the circles,  results in a semi-Dirac point of type I,  with zero Berry phase ($\Delta = 0$). The $\Delta>0$ phase describes a trivial band insulator. Bottom row:  three proximate anisotropic Dirac cones ($\Delta <0$) coalesce into a semi-Dirac point of type II, which has a $\pi$ Berry phase (modulo $2\pi$). The $\Delta>0$ phase is semimetallic, with a single Dirac point.  }
\label{fig:Lifshitz} 
\end{figure}

The annihilation of two Dirac points related by time reversal symmetry produces a semi-Dirac point with  zero Berry phase and winding number. Their merger can be described by a universal Hamiltonian \cite{montambaux2009universal, Adroguer2016},
\begin{equation}
\label{H_Delta}
    \hat{\mathcal{H}}_\Delta(\mathbf{k})=\left(\frac{k_x^2}{2m}+\Delta\right)\sigma_x  + v k_y \sigma_y,
\end{equation} 
that is controlled by the tuning parameter $\Delta$ as illustrated in the top row of Fig. \ref{fig:Lifshitz}. $\sigma_x$ and $\sigma_y$ denote the off-diagonal $2\times2$ Pauli matrices in pseudo-spin space. The low energy spectrum for $\Delta <0$ displays two Dirac points at $\mathbf{k} =  (\pm\sqrt{-2m \Delta},0)$, evolving  to a fully gapped state at $\Delta>0$. At the Lifshitz transition $\Delta=0$ the two bands touch at the semi-Dirac point. The anisotropic quasiparticles around this point have been called type-I semi-Dirac fermions, and propagate with the energy spectrum described in Eq. (\ref{SemiDirac_spectrum}). 
In contrast, the coalescence of two Dirac points with the same winding number results in a parabolic band touching point \cite{Montambaux2018b} with winding number 2 and zero Berry phase (modulo $2\pi$), as in graphene bilayers \cite{CastroNeto2009}. 

Addition of a uniform mass term $\Delta_z \sigma_z$  to Hamiltonian (\ref{H_Delta}) gaps out the semi-Dirac point and produces a finite Berry curvature $\Omega^a_z(\mathbf{k})= \nabla_\mathbf{k} \times i\langle a, \mathbf{k}| \nabla_\mathbf{k} | a, \mathbf{k}\rangle \cdot \hat{\mathbf z}= a \Delta_z v k_x/[2m \varepsilon^3(\mathbf{k})] $ \cite{Saha2016}, where $| a, \mathbf{k}\rangle$ are the Hamiltonian eigenkets  for band $a=\pm$. This quantity is antisymmetric in the momentum coordinates under inversion around the semi-Dirac point, $\Omega^a_z(-\mathbf{k}) = -\Omega^a_z(\mathbf{k})$, and hence produces zero topological charge $\int \mathrm{d}^2k\,\Omega^a_z(\mathbf{k})=0$. This contrasts with both the case of massive Dirac fermions in 2D \cite{Semenoff2008}, and parabolic band touching points with a mass gap \cite{Nandkishore2010,Jung2011}, which have topological charge $\pm\frac{1}{2}$. 

The merger of an odd number of Dirac points produces semi-Dirac fermions of type II, which have a $\pi$ Berry phase and can be topologically non-trivial \cite{VanderbiltTypeII}. The simplest type-II semi-Dirac point results from the merger of  three Dirac cones (see Fig. \ref{fig:Lifshitz}, bottom row). This process is  described by the Hamiltonian 
\begin{equation}
\label{typeII_H}
    \hat{\mathcal{H}}_{\Delta,\text{II}}(\mathbf{k})=\left(\frac{k_x^2}{2m}-vk_y+\Delta\right)\sigma_x+ \frac{ k_x k_y}{m_2} \sigma_y, 
\end{equation} 
where the constant $m_2$  has units of mass. $\Delta$ controls a topological transition between a gapless phase ($\Delta <0$), with three Dirac cones at coordinates $\mathbf{k}= (0,\Delta/v), (\pm\sqrt{-2m\Delta},0)$ and total Berry phase $\phi=\pi$ (modulo $2\pi$),  and another gapless phase ($\Delta>0$) with a single Dirac point at $\mathbf{k} = (0,\Delta/v)$. 
The latter phase ($\Delta >0$) hosts fermionic Dirac excitations at the lowest energies in the vicinity of the nodal point, crossing over to type-II semi-Dirac behavior with increasing energy.  
The phase boundary of the topological Lifshitz transition at $\Delta=0$ exhibits a type-II semi-Dirac point, which retains the total non-trivial Berry phase of the three merged Dirac points, as shown in Fig. \ref{fig:Lifshitz}. Semi-Dirac quasiparticles of this type have the energy spectrum
\begin{equation}
\varepsilon_{\pm}(\mathbf{k}) = \pm\sqrt{(k_x^2/2m - v k_y)^2 +  k_x^2 k_y^2/m_2^2},
\label{EII}
\end{equation}
where $\pm$ label the conduction and valence bands.

Type-II semi-Dirac points acquire a finite topological charge of $\pm\frac{1}{2}$ when gapped out by a mass term $\Delta_z \sigma_z$, and thus are required by topological constraints to appear in pairs in the Brillouin zone \cite{Nielsen1981, Nielsen1981b}. Systems with type-II semi-Dirac points can develop a finite Chern number in the presence of a mass term that breaks global time reversal symmetry \cite{VanderbiltTypeII,Saha2016}. 

Several mechanisms have been proposed to tune across the topological Lifshitz transition described in Eq. (\ref{H_Delta}) and (\ref{typeII_H}), both in solids and in optical lattices of cold atoms. Most theoretical proposals are based on deformed honeycomb lattices \cite{MontambauxMerging, PereiraStrain,montambaux2009universal,Bena2011,wunsch2008dirac,polini2013artificial} and  square lattices under staggered potentials \cite{Delplace2010}, where manipulation of lattice parameters can cause motion, nucleation, and coalescence of Dirac cones. 

Semi-Dirac fermions were theoretically identified as emergent low energy excitations in the extended repulsive Hubbard model on the honeycomb lattice \cite{Christou2018}, where a gapless charge order that triples the length of the unit cell was found in the ground state. 
These quasiparticles were also encountered in the ground state of the anisotropic spin-$\frac{1}{2}$ honeycomb Kitaev model with a transverse Ising field \cite{Hu2024}. 
The quantum critical point separating a gapped quantum spin liquid phase from a gapless one is described by a Lifshitz transition where two Majorana fermion Dirac cones merge. In the same way, a  spin liquid with semi-Dirac Majorana fermions was predicted in the magnetically frustrated trellis lattice \cite{Chaterjee2026}.  Semi-Dirac fermions of type-I and type-II have been theoretically found in twisted bilayers composed of two rotated 2D nodal superconductors \cite{Volkov2023}.     The twist angle controls a sequence of two separate Lifshitz transitions that describe the merger of two and three Dirac cones, respectively. 

\subsection{Density of states} 
\label{sec:DOS}

\begin{figure}
            \centering
              \includegraphics[width=0.48\textwidth]{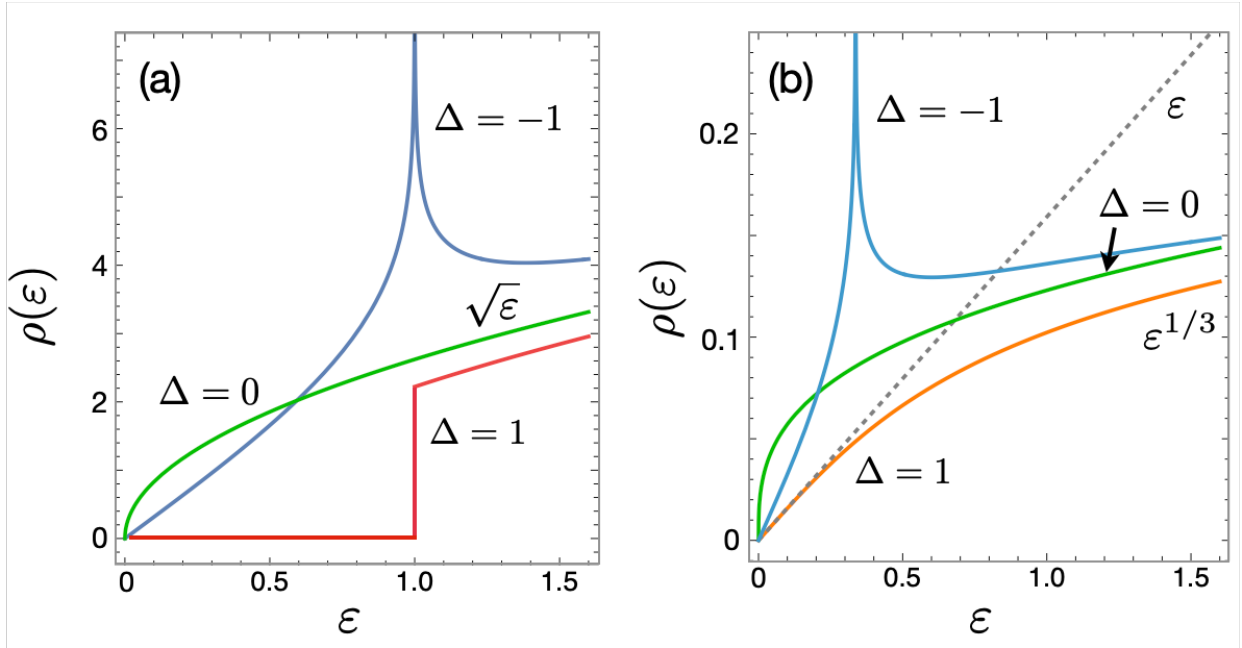}
\caption{ Density of states $\rho(\varepsilon)$  across the Lifshitz transition, in units of $2\pi^2 v/\sqrt{2m}$, versus energy (in units of  $2mv^2$). (a) Type-I semi-Dirac fermions (green curve), with $\rho\propto\sqrt{\varepsilon}$ at $\Delta=0$. Red curve: gapped phase ($\Delta = 1$). Blue curve: Dirac phase ($\Delta=-1$), reflecting a van-Hove singularity located halfway between two merging Dirac cones. (b) Type-II case, where $\rho \propto \varepsilon^{1/3}$ (green curve) at $\Delta=0$.
The value of $m_2$ is fixed so that
$m_2 v^2 = 1$. 
Orange curve: the merger of three Dirac cones through Hamiltonian (\ref{typeII_H}) produces a hybrid semi-metallic phase for $\Delta>0$, where the density of states $\rho(\varepsilon) \propto \varepsilon^\eta $ crosses over from Dirac fermion behavior ($\eta=1$, dashed line) at low energy to type II semi-Dirac ($\eta=\frac{1}{3}$) at higher energy. Blue curve: Dirac phase ($\Delta=-1$).}
\label{fig:DOS_Fig} 
\end{figure}

The anisotropic nature of semi-Dirac quasiparticles is expressed in the sublinear scaling of the density of states with energy. In the semi-Dirac phase $\Delta=0$ of  (\ref{H_Delta}) the density of states can be calculated through the momentum mapping 
$
p_x^2/2m = \varepsilon \cos\theta$ and $vp_y = \varepsilon \sin \theta
$,
with $\theta\in [-\frac{\pi}{2},\frac{\pi}{2}]$ and $\varepsilon>0$, in a way that the dispersion relation (\ref{SemiDirac_spectrum}) is automatically satisfied. This mapping is unambiguously defined in each separate half-plane  $p_x \lessgtr0$ and can be extended across the topological Lifshitz transition for arbitrary $\Delta$ \cite{Adroguer2016}. The density of states is the Jacobian of the transformation $(k_x,k_y) \to (\varepsilon, \theta)$ after integration over the angular variable $\theta$,  \cite{Banerjee2012,Adroguer2016, montambaux2009universal}, 
\begin{equation}
    \rho(\varepsilon)=\frac{1}{2\sqrt{2\pi}\,\Gamma^2(\frac{3}{4})} \frac{\sqrt{2m}}{v} \sqrt{\varepsilon},
    \label{DOS-I}
\end{equation} where $\Gamma(x)$ is a gamma function. In the Dirac phase $\Delta<0$ the low energy density of states is linear in energy, $\rho(\varepsilon)\propto \varepsilon$, with an additional two-fold cone degeneracy. The Dirac cones are strongly anisotropic, connected by a van Hove singularity at the saddle point located half-way between them, along the massive direction of (\ref{H_Delta}), where the density of states diverges logarithmically [Fig. \ref{fig:DOS_Fig}(a)].

 For type-II  semi-Dirac fermions described by Hamiltonian (\ref{typeII_H}), the density of states is \cite{elsayed2026interacting},
\begin{equation}
    \rho(\varepsilon)=\frac{1}{6\pi^{\frac{3}{2}}}\frac{\Gamma(\frac{1}{6})}{ \Gamma(\frac{2}{3}) } \frac{ (2m m_2)^{\frac{1}{3}}  }{ v^{\frac{2}{3} } } \varepsilon^{1/3}.
\end{equation} 
The $\Delta<0$ phase has three strongly anisotropic Dirac cones, with two van Hove singularities located at momentum coordinates
$
(\pm \frac{\sqrt{3}}{2}\sqrt{\gamma_\Delta}v m_2 ,\frac{3}{4}\gamma_\Delta v m_2^2/m),
$  
where $\gamma_\Delta = \sqrt{1-16m\Delta/(9m_2^2 v^2)}-1>0$. As noted earlier, the merger of the three Dirac cones produces a semi-Dirac point at $\Delta=0$, which evolves into a single Dirac cone for $\Delta>0$. The density of states in the latter semi-metallic phase exhibits non-trivial behavior as a function of energy  $\rho(\varepsilon) \propto \varepsilon^{\eta}$, which crosses over from $\eta=1$ scaling at low energy in the vicinity of the Dirac point, to $\eta=\frac{1}{3}$ with increasing energy where it has type II semi-Dirac character [see Fig. \ref{fig:DOS_Fig}(b)]. 

While the density of states of type-I and type-II semi-Dirac fermions, as derived from Hamiltonians (\ref{H_Delta}) and (\ref{typeII_H}), is clearly distinct, the key difference between the two types of quasiparticles is topological. For instance, inclusion of a $t_1 k_x^2 \sigma_0$ term to the type-II Hamiltonian (\ref{typeII_H}) breaks particle-hole symmetry and modifies the energy scaling of the density of states to $\rho(\varepsilon)\propto \sqrt{\varepsilon}$ for finite but small energy $\varepsilon \lesssim  t_1 (2mv)^2$, similarly to the type I case, crossing over to $\rho(\varepsilon)\propto \varepsilon^{1/3}$ at $\varepsilon \gg  t_1 (2mv)^2$. This extra term appears in the effective 2D Hamiltonian of a nodal-line semimetal [see Eq. (\ref{Eq:CP}) in Sec. \ref{sec:nodal-line}], with an external magnetic field used to slice the Brillouin zone near a nodal-line crossing where three Dirac cones merge \cite{BasovZRSIS}.  Fingerprints distinguishing between type-I and type-II quasiparticles are thus more clearly expressed through observables that are sensitive to the Berry phase and other topological invariants. 

Extensions to Hamiltonians of generalized semi-Dirac fermions have been proposed \cite{RoyFoster, Elsayed2025, ElsayedQuartic}, in which the non-relativistic direction has arbitrarily flat bands that disperse with an integer power $k_x^{n}$, with $n\geq1$. In these models, the density of states scales as  $\rho(\varepsilon) \propto \varepsilon^{1/n}$.

\subsection{Spectrum of Landau levels}
\label{sec:LL}

\begin{figure}
            \centering
              \includegraphics[width=0.48\textwidth]{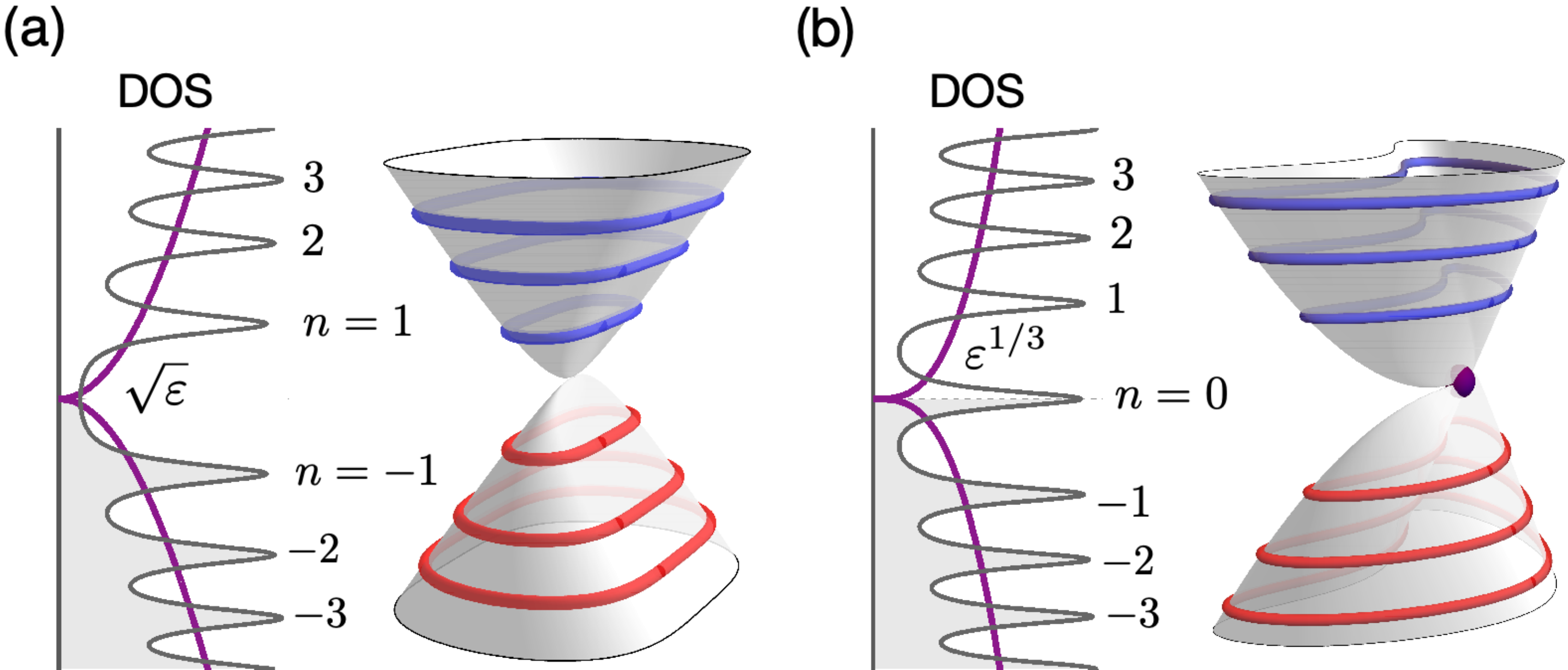}
\caption{ Illustration of the energy scaling of the density of states (DOS) and the LL energy spectrum for  type-I (a) and  type-II (b) semi-Dirac fermions in the presence of a magnetic field $B$. The energy of the levels scales with the signature $B^{2/3}$ and $B^{3/4}$ dependence with magnetic field, respectively. Purple line: zero-field density of states. Type-II semi-Dirac fermions have a $\pi$ Berry phase and show a zero energy LL ($n=0$), in contrast with the type-I case.  }
\label{fig:Bfield} 
\end{figure}

The signature of semi-Dirac fermions under a uniform  magnetic field $B$ is revealed through the spectrum of Landau levels (LL). On phenomenological grounds, the energy spectrum can be determined semiclassically through the quantization of the area enclosed by a fixed energy trajectory in momentum space, $S(\varepsilon)= 4\pi^2\int_0^\varepsilon \mathrm{d}\varepsilon^\prime \rho(\varepsilon^\prime)= 2\pi (|m|+\gamma)eB$, with $m$ an integer and $0\leq \gamma<1$ a phase factor \cite{MontambauxLL,Onsager1952}. After integration over the density of states (\ref{DOS-I}), the energy spectrum  of type-I semi-Dirac fermions at weak field can be easily  found to have the form
\begin{equation}
    \varepsilon_{n}=\mathrm{sgn}(n)\left[\frac{3\Gamma^{2}(\frac{3}{4})}{\sqrt{2\pi}}\,\frac{v}{\sqrt{2m}}(|n|-1+\gamma)eB\right]^{2/3},
\end{equation}
with $n$ a non-zero integer ($|n|\geq 1$), and $\mathrm{sgn}(n)=\pm1$ labeling positive and negative energy states. The phase factor $\gamma = \frac{1}{2} - \frac{\phi}{2\pi}$  is determined from the microscopic Hamiltonian and is set by the Berry phase $\phi$ (modulo $2\pi$) around the semi-Dirac points \cite{MontambauxLL, PardoPickettTB}.  Type I semi-Dirac fermions  thus lack a zero energy LL  ($\gamma =\frac{1}{2}$), in contrast with type II semi-Dirac fermions, which have a $\pi$ Berry phase ($\gamma=0$) [see Fig. \ref{fig:Bfield}], as in conventional Dirac fermions in 2D \cite{CastroNeto2009, Goerbig2011}.
Across the topological Lifshitz transition describing the merging of two Dirac points, the LL spectrum continuously evolves from $\varepsilon_n \propto\sqrt{|n| B}$ in the Dirac semimetal phase ($n\in \mathbb{Z}$) to $\varepsilon_n\propto (|n|-\frac{1}{2})B$ in the gapped one ($|n|\geq 1$), with $\varepsilon_n\propto [(|n|-\frac{1}{2})B]^{2/3}$ in between ($\Delta =0$)  
\cite{Delplace2010}. 

The LL spectrum for type-II semi-Dirac fermions resulting from the merger of three Dirac cones, as described by Eq. (\ref{typeII_H}) at $\Delta=0$,  is
\begin{equation}
    \varepsilon_{n}= \mathrm{sgn}(n) \left[4\sqrt{\pi}\,\frac{\Gamma(\frac{2}{3})}{\Gamma(\frac{1}{6})}\left(\frac{v^2}{2m m_2}\right)^{\frac{1}{3}}|n| e B\right]^{3/4}\,,
\end{equation} 
with $n\in \mathbb{Z}$. At  $\Delta<0$, the LL spectrum has the conventional Dirac form $\varepsilon_n \propto \sqrt{|n|B}$. In the $\Delta >0$ phase it exhibits both Dirac and semi-Dirac behavior  depending on the energy scale, consistently with the behavior of the density of states at zero field, $\rho(\varepsilon)\propto\varepsilon^\eta$, shown in Fig. \ref{fig:DOS_Fig}. 
The spectrum in this phase is  $\varepsilon_n \propto (|n| B)^{1/(\eta +1)}$, crossing over from $\varepsilon_n \propto \sqrt{|n| B}$ scaling at low energy ($\eta=1$) to $\varepsilon_n \propto (|n|B)^{3/4}$ at higher energy ($\eta =\frac{1}{3}$).  

Application of an  in-plane electric field changes the  semiclassical quantization condition of the energy and can produce LL collapse when the corresponding fixed energy trajectory in momentum space becomes an open curve. LL collapse for type I semi-Dirac fermions is highly anisotropic and can occur when a strong electric field, larger than $vB/c$ (in the uniform case), is applied  along the non-relativistic direction \cite{asafov2024}. 

Another distinguishing feature of type-I semi-Dirac materials is their magnetic response to external $B$ fields. The orbital magnetic susceptibility is always diamagnetic and the ratio of orbital to spin susceptibilities at low energies is unusually large  \cite{Banerjee2012}. This contrasts with the behavior of the orbital susceptibility of lightly doped Dirac fermions \cite{Principi2010, Ghosal2007}, which is vanishingly small, and also with the behavior of systems with parabolic bands, where the diamagnetic response is subdued by screening effects \cite{Mahan2000}.

\subsection{Transport}
\label{sec:Transport}

\subsubsection{Optical  conductivity}
\label{sec:Optical}

As expected, the longitudinal charge transport of semi-Dirac fermions is strongly anisotropic  \cite{Carbotte2019,Banerjee2012,Oriekhov2022,Adroguer2016}. In the Dirac phase  the optical conductivity in the relativistic direction $\sigma_{yy}(\omega)$ exhibits a giant response \cite{mawrie2019,Oriekhov2022}, in contrast with  the conductivity  along the direction the Dirac cones merge, $\sigma_{xx}(\omega)$, which is non-resonant. This is due to the symmetry in momentum of the interband matrix elements of the velocity operator,  proximate to the van Hove singularity at the saddle point connecting the nearby cones. Specifically, the matrix elements are symmetric (anti-symmetric) in momentum around the van Hove when the  velocity operator points along the relativistic  (merging) direction \cite{mawrie2019}. 

At the Lifshitz transition $\Delta=0$, the optical conductivity for type-I semi-Dirac fermions is \cite{Carbotte2019, Oriekhov2022}
\begin{align}
\sigma_{xx}(\omega) &=\frac{e^2}{h} \frac{1}{5G}  \sqrt{\frac{\omega}{mv^2}}\theta(\omega -2\varepsilon_F) \label{sigma_xx} \\
\sigma_{yy}(\omega) &=\frac{e^2}{h} \frac{\pi G}{6}  \sqrt{\frac{mv^2}{\omega}}\theta(\omega -2\varepsilon_F),
\label{sigma_yy}
\end{align}
where $\varepsilon_F>0$ is the Fermi energy away from neutrality, $G=\Gamma^2(\frac{1}{4})/2\pi^{\frac{3}{2}}\approx 0.834$ is Gauss's constant and $\theta(x)$ is a step function. The optical response in the clean limit is entirely due to interband contributions. Despite  the strong anisotropy, the geometric mean of the optical conductivities $\sqrt{\sigma_{xx}\sigma_{yy}}=e^2/h(\sqrt{\pi/30})$ is universal  \cite{Carbotte2019}.   The universal geometric mean and the frequency scaling of the conductivities (\ref{sigma_xx}) and (\ref{sigma_yy}) persist in the presence of a finite tilt of the semi-Dirac cone in the relativistic direction, up to a critical tilt \cite{yan2023a}. 
 The analytic derivation of the optical conductivity for type-I semi-Dirac fermions across the topological Lifshitz transition can be found in \cite{Oriekhov2022}.

The Dirac phase in type-II systems exhibits a giant optical conductivity along both principal axes \cite{xiong2023}, reflecting the existence of an additional van Hove singularity in the bands compared to the type-I case.  
The peculiar optical properties and momentum anisotropy of the dielectric function of semi-Dirac fermions manifest in a variety of phenomena related to surface interactions with light. For instance, polarization-selective effects such as linear dichroism \cite{ZhouDichroism}, and  large longitudinal spatial displacement of a light beam upon reflection, known as the  Goos-H\"{a}nchen shift \cite{hong2025goos,xiang2024}, have been predicted. Moreover, semi-Dirac fermions were proposed to mediate a strong photonic spin Hall effect, which is expressed in the separation of reflected left and right circularly polarized light \cite{hong2025goos}.

\subsubsection{DC  conductivity}

 The longitudinal DC conductivity of semi-Dirac fermions at zero temperature, in the clean limit, away from the neutrality point, is $\sigma_{xx}(\omega) \sim (e^2/h) \varepsilon_F^{3/2} \delta(\omega)$, and $ \sigma_{yy}(\omega) \sim (e^2/h) \sqrt{\varepsilon_F}\delta(\omega)$  
 \cite{Carbotte2019}.
In this regime, interband processes are Pauli blocked and transport is due solely  to intraband Fermi surface contributions. At half filling,  the intraband conductivity vanishes and charge transport is dominated by  interband processes. Phenomenological incorporation of a finite scattering rate $\eta$ due to impurities \cite{Carbotte2019} results in a non-universal minimum of conductivity 
 \begin{equation}
 \label{min}
 \sigma_{xx}(0) = \frac{e^2}{h} \frac{1}{\pi G} \sqrt{\frac{\eta}{mv^2}},\quad 
  \sigma_{yy}(0) = \frac{e^2}{h} \frac{G}{2} \sqrt{\frac{mv^2}{\eta}}. 
 \end{equation}
This contrasts with the well known result for Dirac fermions in 2D, where the minimum of conductivity with unitary disorder,  $ \sigma(0) = e^2/(\pi h)$, is isotropic, universal, and independent of the scattering rate \cite{Fradkin1986, Peres2010}. While the general structure of the minimum conductivity (\ref{min}) does not typically depend on how disorder is introduced, numerical prefactors may change when disorder is accounted for self-consistently.

According to Fermi's golden rule, the elastic scattering rate of semi-Dirac fermions due to unitary uncorrelated disorder has a non-trivial angular dependence with momentum \cite{Adroguer2016}. A random scattering potential $V(\mathbf{r})$, such that $\langle V(\mathbf{r}) \rangle=0$ and $\langle V(\mathbf{r}) V(0) \rangle=\gamma \delta(\mathbf{r})$, produces the scattering time $\tau(\varepsilon, \theta)= \hbar/[\pi \gamma \rho(\varepsilon)(1+r \cos\theta)] $, where $r=2\Gamma^4(\frac{3}{4})/\pi^2 \approx0.46$ and $\theta$ is the angular variable introduced before Eq. (\ref{DOS-I}). The DC conductivity in the diffusive regime $\varepsilon_F \gg \hbar \tau^{-1}$ is  \cite{Adroguer2016}
\begin{equation}
\sigma_{xx}(0) \approx 0.20 \frac{e^2\hbar}{\pi \gamma} \frac{2\varepsilon_F}{m},\quad \sigma_{yy}(0) \approx 1.49 \frac{e^2\hbar}{\pi \gamma} v^2.
\end{equation}
This result agrees qualitatively  with the semiclassical conductivity \cite{Banerjee2012}, obtained through averaging the square of the quasiparticle velocity around the anisotropic Fermi surface.  The conductivity along the relativistic direction is independent of the Fermi energy, in contrast with the conductivity of Dirac fermions in the diffusive regime, $\sigma(0)= \frac{1}{2}(e^2/h) v_F^2 \rho(\varepsilon_F)\tau(\varepsilon_F) $  \cite{Peres2010}, which is proportional to the density of states.
Disorder can drive a topological Lifshitz transition from an insulating phase to a semimetal \cite{Sriluckshmy2018}. 

As in graphene, semi-Dirac fermions undergo Klein tunneling when scattered through an arbitrarily large potential barrier \cite{Ghasemian_2021}.  For normal incidence, semi-Dirac fermions have complete transmission  along the massless direction and total reflection along the massive direction, with smooth interpolation between the two limits at intermediate angles \cite{Banerjee2012}. Thermal and thermoelectric transport for type-I semi-Dirac fermions has been addressed in  \onlinecite{mawrie2019thermo, Mandal2020, Mandal2024, vargiamidis2025}. A review on transport of anisotropic quasiparticles, including semi-Dirac fermions, can be found in \onlinecite{Rudenko_2024}.

\subsubsection{Hall conductivities}
\label{sec:Hall}

The  linear Hall conductivity   $\sigma_{xy}(\omega)$ is generically required to be zero by Onsager's reciprocity relations when time-reversal symmetry is present \cite{Sodemann2015}, and can be non-zero when that symmetry is broken  \cite{Xiao2010}. As in any system with a finite Hall conductivity, semi-Dirac fermions can exhibit the Kerr (Faraday) effect, in which the axis of polarization of linearly polarized light that is reflected from (transmitted through) the sample rotates. The Kerr (Faraday) angle in 2D is $\theta_\mathrm{R}(\omega) \sim 2\pi \mathrm{Re}\sigma_{xy}(\omega)/c$, times a factor of order 1 that depends on the index of refraction of the system,  with $c$ the speed of light \cite{Nandkishore2011, SinhaNanoribbon}. In the Chern insulator phase, where $\sigma_{xy}(0)$ is finite and quantized by the Chern number in units of $e^2/h$, the Kerr (Faraday) rotation in the DC limit is of order of the quantum electrodynamics (QED) fine structure constant, $\theta_\mathrm{R}(0) \sim e^2/(\hbar c) = 1/137$, which corresponds to a rotation of a few degrees.

\begin{figure}
           \centering
               \includegraphics[width=0.48\textwidth]{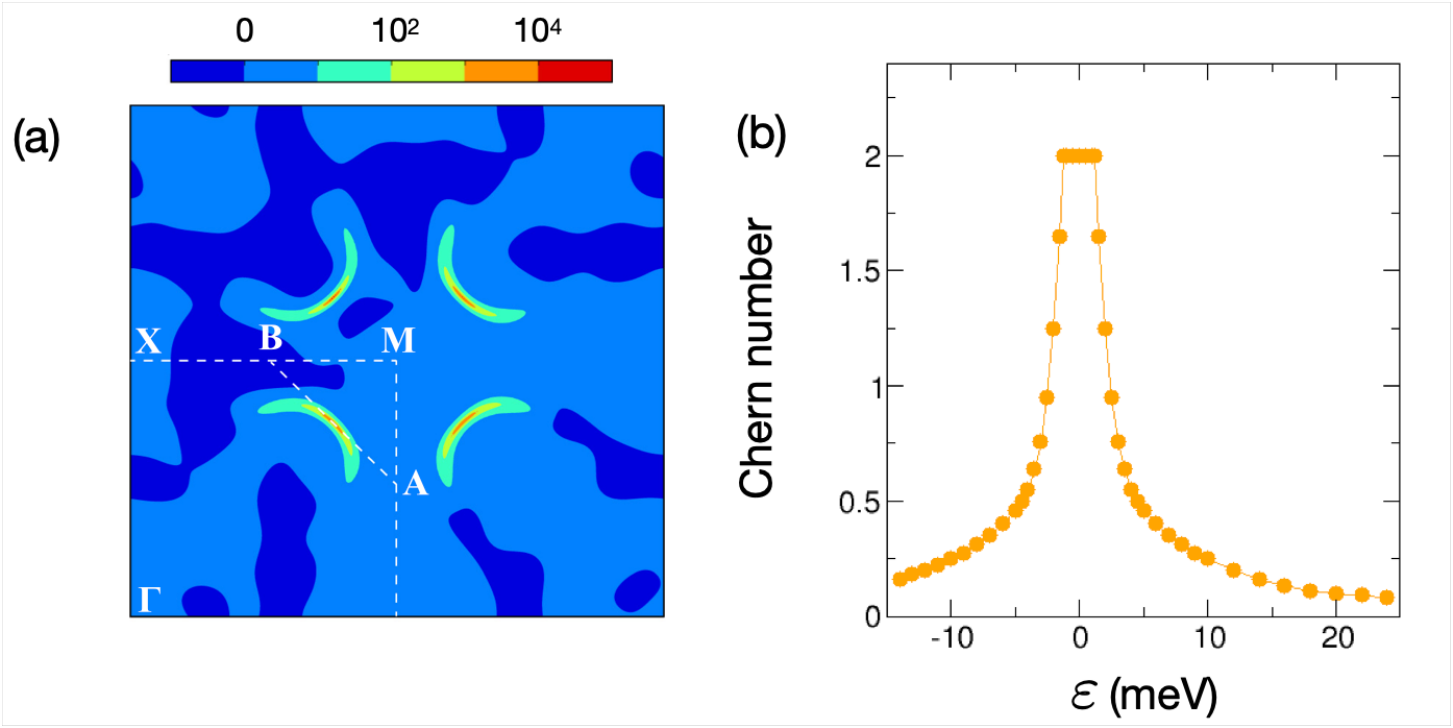}
\caption{  (a)  Density plot of the Berry curvature $\Omega_z(\mathbf{k})$ throughout the Brillouin zone of a TiO$_2$/VO$_2$ heterostructure, as derived by {\it ab initio} methods, in the  presence of spin-orbit coupling. Gapped type-II semi-Dirac points form at the four red areas with Berry curvature accumulation. (b) Predicted total Chern number of  TiO$_2$/VO$_2$ as a function of energy away from the Fermi level. Chern number is quantized inside the spin-orbit coupling gap. Adapted from \onlinecite{VanderbiltTypeII}.  }
\label{fig:Berry curvature} 
\end{figure}

Chern bands can be realized in systems with semi-Dirac fermions either through penetration of a magnetic flux  in the quantum Hall regime,  with irradiation of circularly polarized light, or else  through hopping and interactions. Application of circularly polarized light in gapless semi-Dirac fermion bands of either type I  \cite{Saha2016} or type II \cite{chen2022a} can lead to emergent  Chern insulating phases.  
Chern bands were predicted to emerge spontaneously in TiO$_2$/VO$_2$ heterostructures in the presence of spin-orbit coupling (see Fig. \ref{fig:Berry curvature}) which gaps out a set of four type-II semi-Dirac points that form in the Brillouin zone  away from high symmetry points \cite{VanderbiltTypeII}. Chern bands were also found in lattice models with extended complex hopping terms  \cite{Mondal2022}. 
The chiral edge modes of semi-Dirac fermion Chern insulators  can be unusual, and may have cubic rather than linear dispersion \cite{Olmos2026}.
The zero frequency Hall conductivity of type-I semi-Dirac fermions due to a magnetic flux is quantized in units of $2e^2/h$ per valley \cite{ZhouDichroism, SinhaNanoribbon}.   

Besides the linear Hall response, systems with semi-Dirac quasiparticles can have unconventional non-linear Hall response to external electric fields. This is due to the fact that the Berry curvature $\Omega_z^a(\mathbf{k})$ of  semi-Dirac fermions with a mass gap  breaks inversion symmetry around the semi-Dirac point, as discussed below Eq. (\ref{H_Delta}),  even in the absence of spin-orbit coupling or warping effects \cite{Samal2021}. The broken spatial symmetry is a consequence of the merger of Dirac cones at the topological Lifshitz transition. This effect can produce a finite Berry curvature dipole 
\begin{equation}
\mathbf{D} =  \sum_a \int_\mathrm{BZ} \frac{\mathrm{d}^2k}{(2\pi)^2}  f[\varepsilon_a(\mathbf{k})] \nabla_\mathbf{k} \Omega^a_z ,
\label{D}
\end{equation}
where $f(\varepsilon_a)$ is the equilibrium Fermi distribution of band $a$, provided the system does not globally preserve the combination of parity and time reversal symmetry 
  \cite{Sodemann2015, Du2018}.  The Berry curvature dipole  defines the non-linear Hall response for the photoconductivity  \cite{Sodemann2015}
\begin{equation} 
\chi_{ijk}(\omega) = -\epsilon_{ik} D_j  \frac{e^3 \tau}{1+i\omega \tau },
\label{chi}
\end{equation}
where $\epsilon_{ij}$ is the antisymmetric tensor ($i,j,k=x,y$) and $\tau$ is the scattering time.  The associated photocurrent is $j_i(\omega) = \sum_{jk} \chi_{ijk}(\omega) \mathcal{E}_j\mathcal{E}_k^*$, where $\boldsymbol{\mathcal{E}}$ is the amplitude vector of the electric field $\mathbf{E}(t)=\mathrm{Re}(\boldsymbol{\mathcal{E}} \mathrm{e}^{i\omega t})$ due to the applied light. 

\begin{figure}
           \centering
               \includegraphics[width=0.48\textwidth]{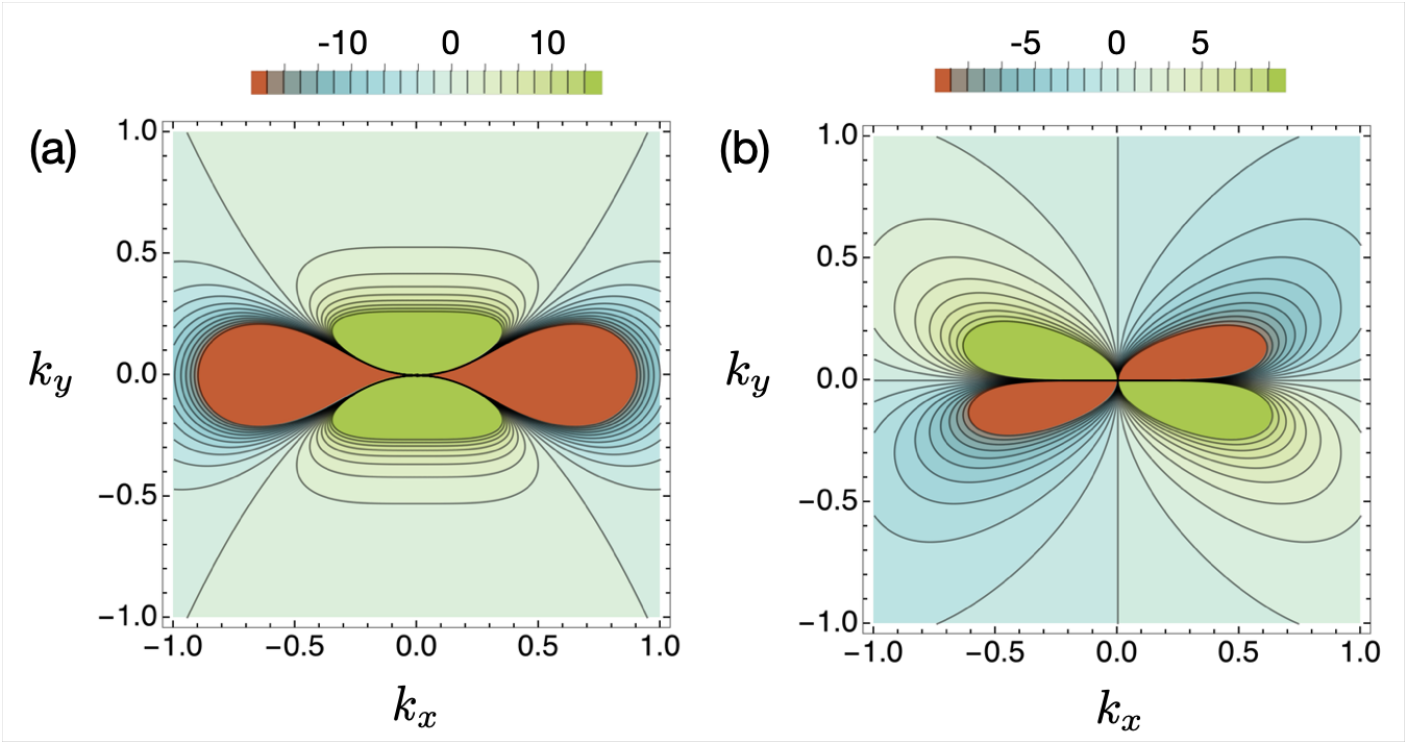}
\caption{(a) Contour plot for the gradient of the Berry curvature of type-I semi-Dirac fermions, $\nabla_\mathbf{k} \Omega^a_z(k_x,k_y)$ along the massive direction, where two Dirac cones merge, and  (b) along the relativistic direction, in the vicinity of the semi-Dirac point. Both panels are symmetric under momentum inversion $\mathbf{k}\to -\mathbf{k}$, but the former (latter) is symmetric (antisymmetric) under $k_x\to-k_x$ or $k_y\to-k_y$. In the presence of a  Fermi surface, gapped semi-Dirac fermions have a finite  Berry curvature dipole oriented along the massive direction, giving rise to the non-linear Hall effect. The Berry curvature dipole along the relativistic direction is zero. Adapted from \onlinecite{Samal2021}.  }
\label{fig:Berry dipole} 
\end{figure}

In the presence of a Fermi surface, type-I semi-Dirac cones with a uniform mass gap $\Delta \sigma_z$ produce  a finite Berry curvature dipole oriented along the massive direction, $D_x \sim -  \Delta (\varepsilon_F -\Delta)^{3/4} /(\sqrt{2m} \varepsilon_F^3) $ \cite{Samal2021}. In the relativistic direction,  $D_y =0$. The orientation of the Berry curvature dipole (\ref{D}) follows from the symmetry properties of the gradient of the Berry curvature $\nabla_\mathbf{k} \Omega^a_z$ near the semi-Dirac point, as shown in Fig. \ref{fig:Berry dipole}. Systems with an odd number of gapped semi-Dirac cones in the Brillouin zone lack inversion symmetry and  hence are required to have a finite Berry curvature dipole when the Fermi level is outside of the gap. Those systems can exhibit a finite photocurrent when pumped with  light. 

Type-II semi-Dirac cones with a mass gap can have a finite Berry curvature dipole (per cone) in both the relativistic and non-relativistic directions \cite{liao2025}. Because of topological constraints, which force them to appear in pairs in the Brillouin zone, the presence of a finite net Berry curvature dipole in the type-II case thus requires additional warping effects that break inversion symmetry globally.

\section{Experimental observations}
\label{sec:experiment}

\begin{figure*}[!ht] 
    \centering
    \includegraphics[width=0.93\textwidth]{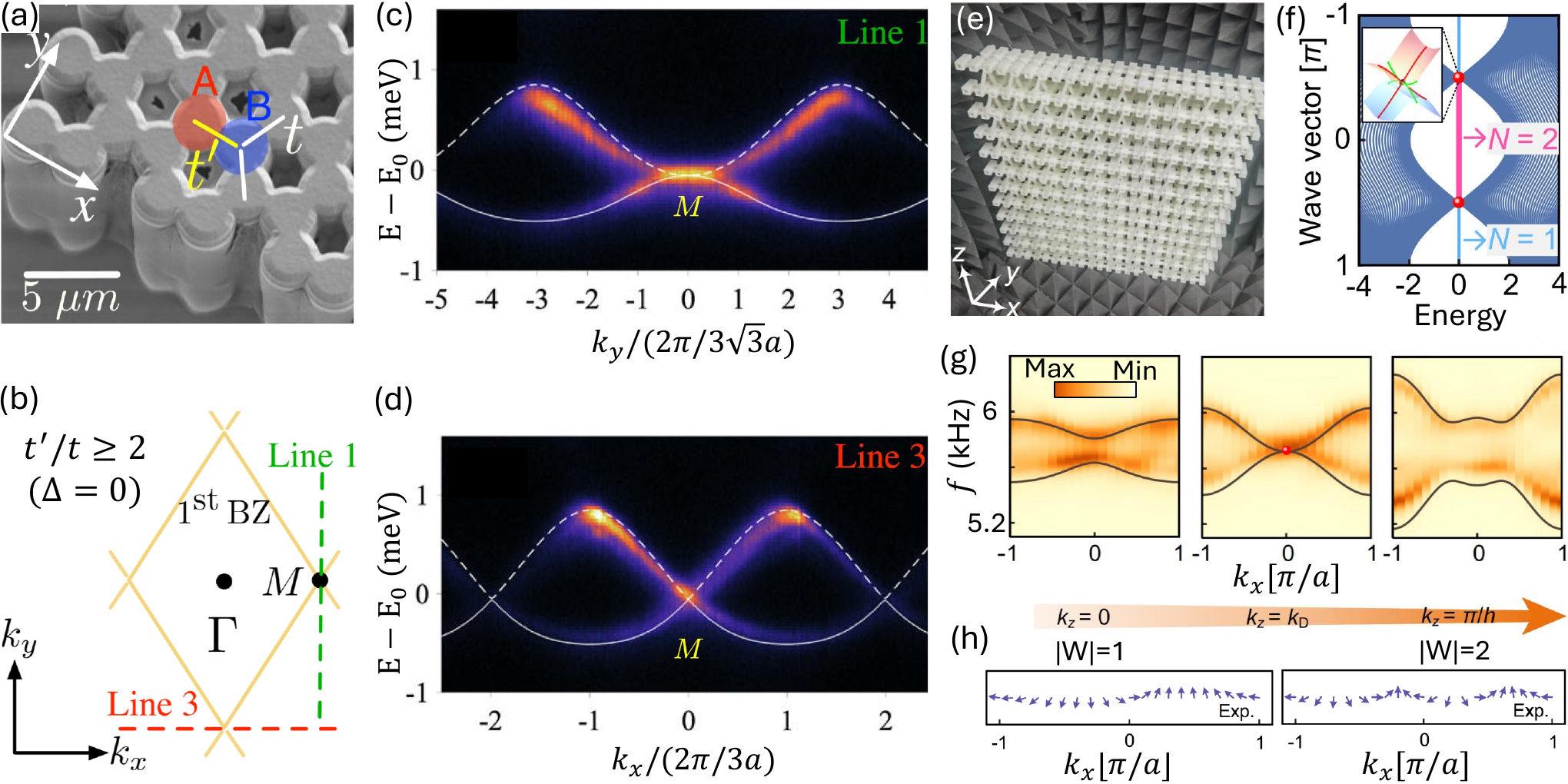} 
    \caption{\label{fig:semidirac_winding}
    Semi-Dirac points, anisotropic dispersion, and winding textures in artificial polariton and sonic lattices. (a) Scanning electron micrograph of a polariton honeycomb lattice; the two sublattices $A$ and $B$ and the anisotropic nearest-neighbor hoppings $t$ and $t'$ are indicated. (b) First Brillouin zone of the compressed lattice ($\alpha=t'/t \ge 2$), showing the momentum cuts denoted line 1 and line 3, both passing through the type-I semi-Dirac point $M$. (c) Momentum-resolved polariton photoluminescence measured along line 1, revealing the quadratic dispersion at $M$. (d) Momentum-resolved photoluminescence measured along the orthogonal cut line 3, revealing the linear dispersion through the same point; in (c) and (d), solid and dashed curves denote tight-binding fits to the lower and upper bands, respectively. (e) Photograph of the sonic crystal used in the acoustic measurements. (f) Projected band diagram of the sonic topological semimetal; blue and red branches denote $N=1$ and $N=2$ zero-energy edge-state pairs, separated by type-II semi-Dirac points (red dots; see inset). In the fabricated crystal, the $N=|W|=2$ interval lies at $|k_z|>k_D$.  (g) Measured bulk bands (color map) versus $k_x$ at $k_z=0$, $k_D$, and $\pi/h$, with tight-binding bands (black curves), where $h$ is the interlayer spacing. The gap closes at $k_z=k_D$ with a quadratic dispersion along $k_x$ (red dot). (h) Normalized Bloch vector $\hat{\mathbf d}(k_x)=(\hat d_x,\hat d_y)$ as a function of $k_x$ reconstructed from the measurements at $k_z=0$ ($|w|=1$) and $k_z=\pi/h$ ($|w|=2$). The Bloch vector has double winding 
    over the Brillouin zone in the topological phase adjacent to type-II semi-Dirac point. Panels (a)-(d) are adapted from Ref.~\cite{Real2020}; panels (e)-(h) are adapted from Ref.~\cite{xiong2025}.} 
    \label{FigS3_syntheticlattice} 
\end{figure*}

\subsection{Synthetic lattices}
\label{sec:synthetic}

Monolayer graphene is a natural candidate to realize type-I semi-Dirac fermions via merging two Dirac cones through tensile strain. However, realizing the Dirac-point merging scenario by straining graphene is experimentally challenging. Tight-binding calculations show that merging the two inequivalent Dirac cones in graphene requires uniaxial tensile strain in excess of $20\%$ ~\cite{MontambauxMerging, PereiraStrain}. Yet, experimental graphene samples fracture at around 6$\%$ tensile strain level ~\cite{cao2020}, well below the threshold for merging Dirac points. Synthetic honeycomb lattices bypass this mechanical limitation because the effective hopping ratio between  anisotropic nearest neighbors, $\alpha=t'/t$, can be tuned directly through geometry or light-matter coupling strength, allowing the system to traverse the topological Lifshitz transition illustrated in Fig. \ref{fig:Lifshitz}. 
A review on the formation of semi-Dirac fermions in synthetic lattices can be found in \cite{Montambaux2018}.

Synthetic lattice semi-Dirac points were first experimentally reported in a Fermi gas of cold atoms on a tunable honeycomb optical lattice \cite{tarruell2012creating}. By varying the intensities of the lattice laser beams, the lattice anisotropy, and hence the effective hopping anisotropy, were tuned, steering the two Dirac cones toward the M point of the Brillouin zone, where they merged and annihilated. Momentum-resolved interband transitions directly tracked the motion of the Dirac points across the topological transition \cite{tarruell2012creating, Lim2012}. The Dirac-point merging transition was subsequently observed in artificial honeycomb platforms, including microwave arrays of dielectric resonators~\cite{Bellec2013} and evanescently coupled photonic waveguide arrays~\cite{Rechtsman2013}. In these systems, the Dirac point merger appeared as strain-driven Dirac-point annihilation at the Brillouin zone boundary, accompanied by gap opening and edge-state reconfiguration.

A direct spectroscopic fingerprint of the anisotropic semi-Dirac dispersion was later obtained from polariton photoluminescence in a honeycomb lattice of GaAs micropillars~\cite{Real2020}, shown in Fig.~\ref{FigS3_syntheticlattice}(a)-(d). At the critical anisotropy $\alpha_c\approx 2$, momentum-resolved spectra along two orthogonal cuts through the M point reveal the hallmark coexistence of a quadratic band touching along line~1 [Fig.~\ref{FigS3_syntheticlattice}(c)] and a linear dispersion along the perpendicular cut [Fig.~\ref{FigS3_syntheticlattice}(d)], in quantitative agreement with tight-binding fits. The same platform also exhibited strongly anisotropic transport and defect-induced localization expected for semi-Dirac quasiparticles.

The preceding synthetic-lattice systems realized type-I semi-Dirac points through the merger of two inequivalent Dirac cones.  Type-II semi-Dirac points require the merger of three Dirac cones and have been realized recently in a sonic crystal of stacked Su-Schrieffer-Heeger (SSH) chains with long-range, intralayer and interlayer couplings~\cite{xiong2025} [see Fig.~\ref{FigS3_syntheticlattice}(e)]. Each momentum slice $k_z$ realizes an extended SSH chain along the $x$ direction. When the intra-to-interlayer coupling ratio reaches a critical value, the bulk gap closes at a pair of type-II semi-Dirac points located at $k_z=\pm k_D$ [Fig.~\ref{FigS3_syntheticlattice}(f)]. Bulk bands  obtained from the measured acoustic pressure field show the gap closing at $k_z=k_D$, with a quadratic dispersion along $k_x$, and reopening at larger $k_z$ [Fig.~\ref{FigS3_syntheticlattice}(g)]. The effective low energy Hamiltonian $\hat{\mathcal{H}}(\mathbf{k}) = {\mathbf d}(\mathbf{k}) \cdot \vec{\sigma}$ 
was obtained through reconstruction of the normalized Bloch pseudospin vector $\hat{\mathbf d}(k_x)$ 
from acoustic transmission data. The pseudo-spin vector has winding number $|w|=1$ at $k_z=0$ and $|w|=2$ at the edge of the Brillouin zone, $k_z=\pi/h$ [Fig.~\ref{FigS3_syntheticlattice}(h)], with $h$ the interlayer distance. The winding number sets the number of zero-energy edge-state pairs  $N=|w|$, as shown in  Fig.~\ref{FigS3_syntheticlattice}(f). Because $w$ changes by one across each semi-Dirac point, the point itself carries winding number 1 and Berry phase $\pi$, as expected for type-II semi-Dirac fermions described in Eq.~(\ref{typeII_H}).

\subsection{Semi-Dirac fermions in solids}
\label{sec:solids}

\begin{figure*}[!ht]
  \centering
  \includegraphics[width=0.96\textwidth]{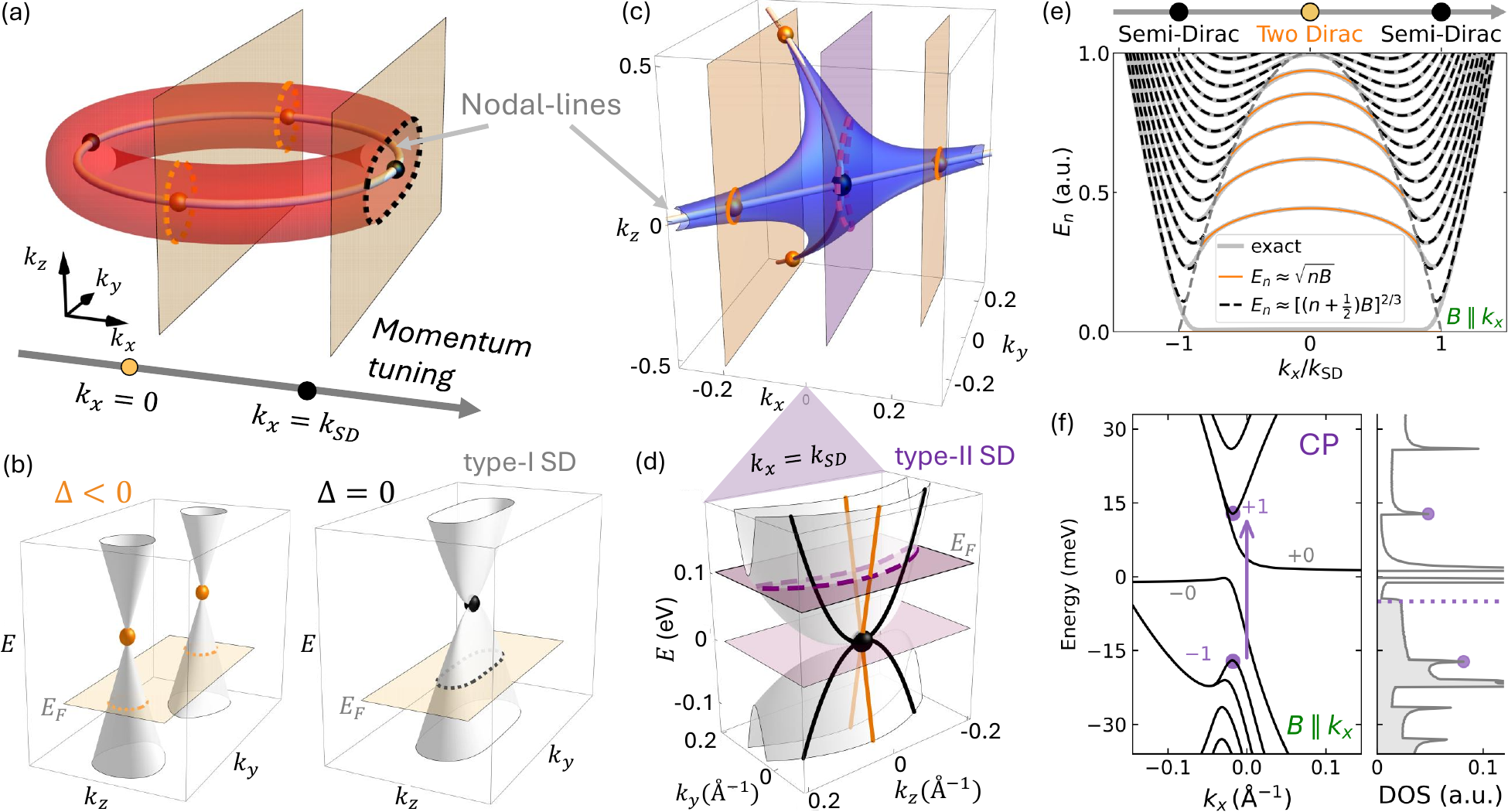}
  \caption{Momentum tuning of Dirac and semi-Dirac physics in nodal-line materials.
    (a) Schematic of a nodal-ring Fermi surface (red) in which changing the momentum slice tunes the effective low-energy dispersion from two Dirac crossings (orange dots) at generic $k_x$ to a semi-Dirac point (black dots) at the critical cut $k_x=k_{\mathrm{SD}}$. The gray arrow indicates momentum tuning of the 2D Hamiltonian (Eq.~\ref{Eq:slice}) along $k_x$.
    (b) Comparison between the band dispersion of two symmetry-related Dirac cones (left) and their merger into a type-I semi-Dirac point at the critical momentum slice (right). Orange shaded planes indicate the Fermi energy $\varepsilon=\varepsilon_F$.
    (c) Schematic of the Fermi surface (blue) of the crossing of two nodal-lines, illustrating the constant-$k_x$ cuts used to access the Dirac (orange shaded planes) and semi-Dirac (purple shaded plane) regimes.
    (d) Type-II semi-Dirac band dispersion at $k_x=k_{\mathrm{SD}}=0$, where the black and orange lines indicate quadratic and linear dispersion along $k_z$ and $k_y$, respectively. The two purple shaded planes indicate two different Fermi levels.
    (e) Landau level (LL) spectrum for magnetic field applied parallel to $\hat k_x$, showing the crossover between the two-Dirac regime and the type-I semi-Dirac regime. The exact spectrum (gray lines) is compared with the semiclassical asymptotic forms for the Dirac regime $\varepsilon_n \approx \sqrt{nB}$ and the semi-Dirac regime $\varepsilon_n \approx [(n+1/2)B]^{2/3}$, where $n$ is the LL index and $B$ is the external magnetic field.
    (f) LL spectrum (left) and density of states (DOS, right) in ZrSiS for $B \parallel \hat k_x$, highlighting the characteristic semi-Dirac quantization and optical transition (vertical purple arrow) between the LLs. Purple dots highlight the singularities in the DOS that dominate the optical transitions. Panel (e) is adapted from Ref.~\cite{oroszlany2018}. Panels (c), (d), and (f) are adapted from Ref.~\cite{BasovZRSIS}.
    }
  \label{FigS3_LLtheory}
\end{figure*}

The synthetic-lattice realizations above establish the essential phenomenology of semi-Dirac points, including the coexistence of linear and quadratic dispersion for type-I semi-Dirac points and the change of winding texture across type-II semi-Dirac points. However, these analogue platforms are limited to systems with  
cold atoms, polaritons, photons, and sound waves. Accessing  nontrivial electronic properties predicted for semi-Dirac fermions, therefore, requires  realizing actual electronic band structures in crystalline solids. 

Semi-Dirac fermions have been theoretically predicted to exist in different quantum materials, including $\alpha$-(BEDT-TTF)$_2$I$_3$ crystals under pressure \cite{katayama2006pressure}, quantum-confined VO$_2$/TiO$_2$ heterostructures \cite{PardoPickettTiO2,VanderbiltTypeII}, transition-metal oxide perovskites \cite{mohanta2021}, and strained crystals such as graphene  \cite{PereiraStrain,amorim2016novel,Bena2011,damljanovic2017existence}, graphene bilayers \cite{deGail2012} and black phosphorus \cite{Rodin2014, rudenko2015}. In VO$_2$/TiO$_2$ heterostructures, the predicted semi-Dirac phase requires precise control of the VO$_2$ thickness \cite{pardo2010} and sharp oxide interfaces. These requirements have made experimental realization challenging: TiO$_2$/VO$_2$ superlattices and ultrathin VO$_2$ films on TiO$_2$ instead showed insulating low-temperature states, persistent VO$_2$-like structural transitions, and interfacial $(\mathrm{V},\mathrm{Ti})\mathrm{O}_2$ formation \cite{Shibuya2010,Muraoka2011}. Angle-resolved photoemission measurements on strongly doped black phosphorus~\cite{Kim2015,Kim2017} provided early indications of semi-Dirac-like electronic structure in a non-synthetic material platform. Nevertheless,  black phosphorus is a semiconductor and requires strong doping to reach the gapless regime.  The precise semi-Dirac dispersion of black phosphorus was not clearly established  experimentally~\cite{Kim2017} or theoretically~\cite{rudenko2015}. 

The first conclusive experimental observation of semi-Dirac fermions in solids was realized in a 3D nodal-line semimetal \cite{BasovZRSIS}.  
These materials offer an unexpected route to 2D semi-Dirac physics, avoiding some of the materials constraints faced by graphene, black phosphorus and oxide heterostructures.

\subsubsection{Momentum-tuned semi-Dirac fermions in nodal-line metals}
\label{sec:nodal-line}

Topological nodal-line semimetals are characterized by band crossings that extend along one-dimensional lines or loops in momentum space rather than occurring at isolated points~\cite{burkov2011,Mullen2015,fang2015, Kim2015b}. This extended nodal geometry in momentum space gives rise to strongly anisotropic low-energy quasiparticles, with linear band dispersion transverse to the nodal line and weaker or vanishing dispersion along it~\cite{ahn2017a,shao2019}. The nodal line geometry and strong anisotropy support characteristic phenomena such as drumhead surface states and unusual bulk optical responses~\cite{armitage2018,lv2021}. Moreover, in real materials, the nodal lines are often dispersive in energy along the nodal line direction and cross the Fermi level $\varepsilon_F$, resulting in degeneracy points very close to it ~\cite{shao2020,wyzula2022}. These features make magneto-optical spectroscopy especially powerful in nodal-line systems by probing the anisotropic band structure with inter-Landau-level transitions~\cite{shao2019, akrap2026}. 

Anisotropic Dirac nodal lines and nodal rings in three dimensions open a new route to semi-Dirac physics by allowing multiple Dirac points to merge in momentum space. For 2D synthetic lattices [see Fig.~\ref{FigS3_syntheticlattice}], the effective tuning parameter in the universal Hamiltonian (\ref{H_Delta}), $\Delta \propto \alpha-\alpha_c$. Thus, for a given lattice, $\Delta$ is fixed by the hopping anisotropy $\alpha=t'/t$. Reaching the semi-Dirac point ($\Delta=0$) therefore requires fabricating or tuning lattices whose anisotropy approaches the critical value $\alpha_c$.  In contrast, for a Dirac nodal ring in 3D momentum space, a semi-Dirac point can emerge within a 2D constant-momentum slice of a single material by continuously varying the third momentum component until the critical slice is reached [$\Delta\propto k_x^2-k^2_{\mathrm{SD}}$ in Fig.~\ref{FigS3_LLtheory}(a),(b)]. 
These effective 2D momentum slices represent cross sections of the bulk band structure, with the critical semi-Dirac slices selectively enhanced in magneto-optical spectroscopy. 

Under a magnetic field $B$ pointing along the third momentum direction, the transverse motion becomes Landau quantized while the momentum parallel to the field is conserved.
The dispersion of the spectrum of LLs along the conserved momentum direction permits selection of the momentum slices by tuning the frequency of the light used in the LL spectroscopy, as indicated by the vertical arrow in  Fig.~\ref{FigS3_LLtheory} (f).  
The singularities in the LL density of states~\cite{oroszlany2018} effectively select the critical semi-Dirac slices in magneto-optical spectroscopy, providing a spectroscopic dimensional reduction without physical confinement~\cite{BasovZRSIS}. 

This mechanism can be exemplified through the toy model Hamiltonian for a nodal ring,
\begin{equation}
\hat{\mathcal{H}}_\mathrm{R}(\mathbf{k})=\frac{k_x^2+k_y^2-k_0^2}{2m}\sigma_x+v_z k_z \sigma_z ,
\label{Eq:NR}
\end{equation}
with energy spectrum
\begin{equation}
\varepsilon_{\mathrm{R},\pm}(\mathbf{k})=\pm \sqrt{\left[(k_x^2+k_y^2-k_0^2)/(2m)\right]^2+v_z^2 k_z^2}.
\end{equation}
Hamiltonian (\ref{Eq:NR}) describes a Dirac nodal ring of radius $k_0$ in the $k_z=0$ plane, as illustrated in Fig.~\ref{FigS3_LLtheory}(a). For a generic cut at fixed $k_x$, the effective two-dimensional Hamiltonian in the $(k_y,k_z)$ plane may be written as
\begin{equation}
\hat{\mathcal{H}}_{\mathrm{2D}}(k_y,k_z)=\left(\frac{k_y^2}{2m}+\Delta_1\right)\sigma_x+v_z k_z\sigma_z,
\label{Eq:slice}
\end{equation}
where $\Delta_1 \equiv (k_x^2-k_0^2)/2m$. With this convention, the momentum slice plays the same role as the detuning parameter in the universal Hamiltonian (\ref{H_Delta}). 

For a generic constant-$k_x$ slice with $|k_x|<k_0$, one has $\Delta_1<0$, and the slice intersects the nodal ring at two points, producing two symmetry-related Dirac crossings in the effective $(k_y,k_z)$ dispersion [Fig.~\ref{FigS3_LLtheory}(b), left]. As $|k_x|$ increases, these two Dirac points move toward each other, and at the critical cut $k_x=k_{\mathrm{SD}}=\pm k_0$ one reaches $\Delta_1=0$, where they merge into a single type-I semi-Dirac point [Fig.~\ref{FigS3_LLtheory}(b), right]. Expanding about this critical slice gives
\begin{equation}
\varepsilon_{\mathrm{SD},\pm} (k_y,k_z)\approx \pm \sqrt{\left(k_y^2/2m\right)^2+v_z^2 k_z^2},
\end{equation}
which is quadratic along $k_y$ (the merging direction) and linear along $k_z$. For $|k_x|>k_0$ ($\Delta_1>0$) the constant-$k_x$ slice no longer intersects the nodal ring and the spectrum becomes gapped. The nodal-ring model in Eq.~\ref{Eq:NR} thus provides a solid-state realization of the universal Hamiltonian (\ref{H_Delta}) for type-I semi-Dirac fermions, at least when interaction effects are not dominant. 

Beyond the two-Dirac-point merger realized in a nodal ring, the connectivity of multiple nodal lines in three-dimensional momentum space~\cite{bzdusek2016,yan2018} can generate the three-Dirac-point merging scenario required for the type-II semi-Dirac Hamiltonian (\ref{typeII_H}). An explicit nodal-line crossing-point (CP) model that realizes this three-Dirac-point merging scenario was proposed in ~\onlinecite{BasovZRSIS}.

\begin{figure*}[!ht] 
  \centering
  \includegraphics[width=\textwidth]{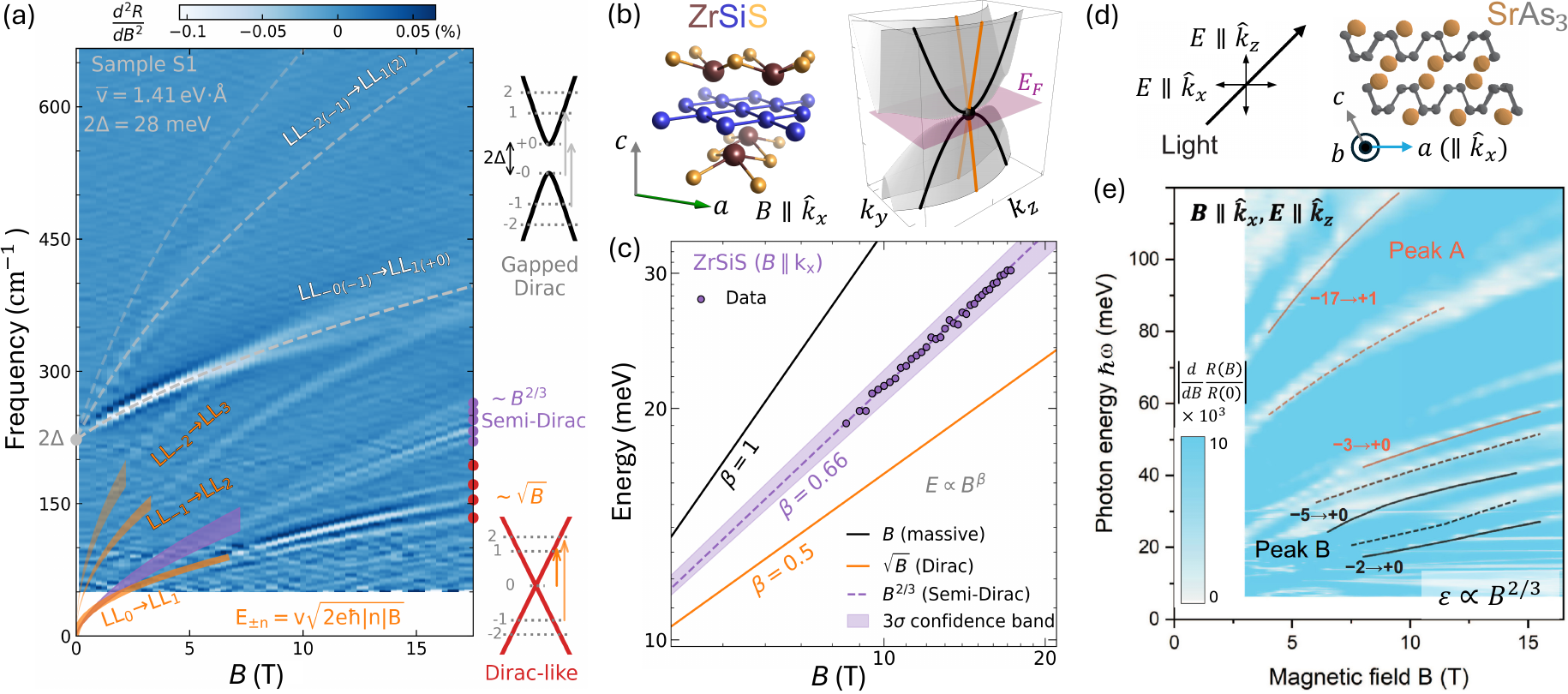}
  \caption{Semi-Dirac fermions identified through Landau-level spectroscopy in nodal-line semimetals ZrSiS and SrAs$_3$.
  (a) Second-derivative magneto-reflectance map of ZrSiS for in-plane magnetic field, showing the main inter-LL transitions across the gapped Dirac bands (right inset) together with weaker subgap branches. The latter separate into Dirac-like transitions with $\sim \sqrt{B}$ and semi-Dirac transitions with $\sim B^{2/3}$ power-laws.
  (b) Crystal structure (left) and band-structure (right) schematics for ZrSiS, illustrating that semi-Dirac fermions arise near a nodal-line crossing where the dispersion is linear along $k_y$ direction and quadratic along $k_z$ direction.
  (c) LL transition energies as a function of $B$ for ZrSiS with ${B} \parallel \hat{k}_x$. Data points (circles) are compared with power-law fits $\varepsilon \sim B^{\beta}$ for $\beta = 0.5$ (Dirac, orange), $\beta = 0.66$ (semi-Dirac, purple dashed), and $\beta = 1$ (massive, black). The shaded region denotes the $3\sigma$ confidence interval.
  (d)~Schematic of the magneto-optical measurement geometry on SrAs$_3$, showing the light polarization ${E}$ relative to the crystal axes.
  (e)~Derivative magneto-reflectance $\frac{d}{dB}\frac{R(B)}{R(0)}$ of SrAs$_3$ with ${B} \parallel \hat{k}_x$ and ${E} \parallel \hat{k}_z$. Two series of LL transitions are labeled Peak~A and Peak~B, with LL indices indicated. Peak B transitions follows $\varepsilon\propto B^{2/3}$, a signature of semi-Dirac fermions. Panels (a)--(c) are adapted from Ref.~\cite{BasovZRSIS}; panels (d), (e) are adapted from Ref.~\cite{jeon2025}.}
  \label{FigS3_LLexperiments}
\end{figure*}

As shown in Fig.~\ref{FigS3_LLtheory}(c), a straight horizontal nodal line and a curved vertical nodal line meet each other at the CP (black sphere). The low-energy electronic structure near this CP is described by the effective two-band Hamiltonian:
\begin{equation}
\hat{\mathcal{H}}_{\mathrm{CP}}
=
 t_1 k_z^2 \sigma_0
+
\left(\frac{k_z^2}{2m}+v k_{\perp}\right)\sigma_z
+
\frac{ k_{\parallel} k_z }{m_2}\sigma_x
+
\delta_\mathrm{so} \sigma_y ,
\label{Eq:CP}
\end{equation}
where $k_{\perp}=k_x-k_y$ and $k_{\parallel}=k_x+k_y$. Here $m$ is the effective mass along $k_z$ and $v$ is the Fermi velocity along $k_{\perp}$.  $t_1$ controls the electron-hole asymmetry along $k_z$, $m_2$ is the effective mass that governs the dispersion along $k_{\parallel}$, and $\delta_\mathrm{so}$ is half of the spin-orbit-coupling  induced gap. Similar to the nodal-ring model (Eq.~\ref{Eq:slice}), the effective mass parameter $\Delta_2 \equiv v (k_x-k_{\mathrm{SD}}) = v k_x $ is tuned by momentum.

Fixing a generic constant-$k_x$ slice, the two-dimensional plane intersects one nodal line once and the other twice, producing three Dirac crossings (orange spheres) in the effective $(k_y,k_z)$ spectrum.  As the slice is tuned toward the critical cut $k_x=k_{\mathrm{SD}}=0$, these three Dirac points approach one another and merge at the nodal-line CP. The resulting band dispersion [Fig.~\ref{FigS3_LLtheory}(d)] is quadratic along $k_z$ and linear along $k_y$, realizing an effective  type-II semi-Dirac fermion Hamiltonian (Eq.~\ref{typeII_H}), ignoring spin-orbit coupling,
\begin{equation}
\hat{\mathcal{H}}_{\mathrm{SD}}(k_y,k_z)=t_1 k_z^2 \sigma_0+
\left(\frac{k_z^2}{2m}-v k_y\right)\sigma_z
+ \frac{k_y k_z}{m_2} \sigma_x .
\label{Eq:SD2}
\end{equation}

The LL spectrum of the nodal-ring model (\ref{Eq:NR}) under a magnetic field ${B}\parallel \hat{k}_x$ is shown in Fig.~\ref{FigS3_LLtheory}(e), where $k_x$ remains a good quantum number labeling the two-dimensional $(k_y,k_z)$ slices normal to the field. The LL spectrum directly encodes the evolution of the effective 2D Hamiltonian with momentum in Fig.~\ref{FigS3_LLtheory}(a): generic $k_x$ cuts show the spectrum for Dirac fermions with $\varepsilon_n\!\sim\!\sqrt{nB}$, while the critical cut $k_x=k_{\mathrm{SD}}$ gives the semi-Dirac fermion scaling $\varepsilon_n\!\sim\![(n+\frac{1}{2})B]^{2/3}$~\cite{oroszlany2018}.
For the nodal-line CP model (\ref{Eq:SD2}), semi-classical analysis predicts that the semi-Dirac LL energy spectrum at the critical cut  scales at low field with $\varepsilon_n \sim (nB)^{2/3}$ for $\varepsilon_n\lesssim t_1(2mv)^2$ (see  discussion in Sec. \ref{sec:DOS}).  
These critical momentum slices correspond to singular behavior in the LL density of states, expected for both type-I semi-Dirac fermions in a nodal-ring~\cite{oroszlany2018} and type-II semi-Dirac fermions in nodal-line CPs~\cite{BasovZRSIS}.  
As we discuss next, LL spectroscopy provides direct experimental access to these singularities through their distinct optical transitions [Fig.~\ref{FigS3_LLtheory}(f)].

\subsubsection{Experimental observation in ZrSiS and SrAs$_3$}
\label{sec:ZrSiS}

Recent magneto-optical experiments on ZrSiS provided the first unambiguous evidence for semi-Dirac fermions in a crystalline material~\cite{BasovZRSIS}. ZrSiS is a layered topological nodal-line semimetal built from a square-net Si layer sandwiched by Zr-S layers above and below~\cite{schoop2016}. This crystal structure gives rise to an intricate three-dimensional nodal-line cage, consisting of two planar nodal squares linked by vertical nodal lines. Rather than originating from an isolated nodal ring, the relevant semi-Dirac fermions in ZrSiS emerge near crossing points of these nodal lines. In particular, the observed low-energy magneto-optical response is dominated by the crossing point in the $k_z=0$ plane [Fig.~\ref{FigS3_LLtheory}(c)], which lies very close to the Fermi level~\cite{BasovZRSIS}. As discussed above, under an in-plane magnetic field $B \parallel \hat{k}_x$, the critical momentum near the nodal-line crossing produces singular features in the $k_x$-resolved LL spectrum and density of states [Fig.~\ref{FigS3_LLtheory}(f)].

The experimental evidence is summarized in Fig.~\ref{FigS3_LLexperiments}(a)--(c). The second-derivative magneto-reflectance map in Fig.~\ref{FigS3_LLexperiments}(a) shows the main inter-LL transitions across the gapped Dirac bands together with weaker subgap branches. These subgap features separate into two distinct classes: Dirac-like branches that scale approximately as $\sqrt{B}$ and semi-Dirac branches that scale as $B^{2/3}$. The latter are the main signature of type-II semi-Dirac quantization in the low field regime $\varepsilon_n\lesssim  t_1 (2mv)^2$, expected at the critical momentum slice. The DFT extracted parameters of ZrSiS for Hamiltonian (\ref{Eq:SD2}) give a crossover energy scale of $t_1 (2mv)^2\approx133$meV \cite{BasovZRSIS}, which is one order of magnitude larger than the energy of the $n=1$ LL ($\varepsilon_1 \approx 15$meV).  
This interpretation is corroborated by the comparison in Fig.~\ref{FigS3_LLexperiments}(c), where the measured transition energies are fitted by power laws $\varepsilon\propto B^\beta$. The extracted exponent $\beta\simeq 0.66$ agrees with the semi-Dirac expectation $\beta=2/3$ 
and is clearly distinct from the Dirac value $\beta=1/2$ and the massive-band value $\beta=1$. Together with the LL calculation based on DFT band structures of ZrSiS in Fig.~\ref{FigS3_LLtheory}(f), these results establish that the observed low-energy excitations originate from the type-II semi-Dirac dispersion near the crossing point of the two nodal lines.

A similar magneto-reflectance experiment later revealed semi-Dirac signatures in the nodal-line semimetal SrAs$_3$~\cite{jeon2025}. In contrast to ZrSiS, the nodal-line structure in SrAs$_3$ is close to a nodal ring and therefore realizes the type-I semi-Dirac scenario (Fig.~\ref{FigS3_LLtheory}(a),(b)) for the merger of two Dirac points in a momentum slice. The experimental geometry is shown in Fig.~\ref{FigS3_LLexperiments}(d). As seen in Fig.~\ref{FigS3_LLexperiments}(e), two families of LL transitions are resolved, labeled Peak~A and Peak~B. Among them, the Peak~B series follows the characteristic $\varepsilon\propto B^{2/3}$ scaling expected for type-I semi-Dirac Landau quantization, providing evidence that the same momentum-slice mechanism anticipated for a nodal-ring system can also be accessed spectroscopically in a real material. Thus, taken together, ZrSiS and SrAs$_3$ furnish complementary experimental realizations of the two solid-state routes to semi-Dirac fermions: the type-II case generated by the crossing of two nodal lines, and the type-I case generated by critical slices through a nodal ring.

\section{Coulomb interactions}
\label{sec:eeinteractions}

\subsection{Perturbation theory}
\label{sec:pert}

As in graphene, the Coulomb interaction for electrons in  2+1 dimensions is mediated by 3D photons and has $1/q$ form with momentum \cite{Gonzalez1994, kotov2012}.  The Coulomb interaction is non-analytic and thus the electron charge is not renormalized \cite{Ye1998,Herbut2006}. Nevertheless, type-I semi-Dirac fermions have a surprisingly unconventional structure of perturbation theory. From now on, unless otherwise noted, we will refer to these quasiparticles simply as semi-Dirac fermions. 

Unlike the standard case of 2D Dirac fermions (graphene) where the velocity is logarithmically renormalized near the Dirac point \cite{Gonzalez1994, kotov2012}, the first order self-energy correction to the  effective inverse mass $m^{-1}$ of semi-Dirac fermions has an unusual $\log^2(\Lambda/\varepsilon)$ divergence in the infrared, with $\Lambda$ the ultraviolet cutoff \cite{ChubukovInteractions, KotovInteractions}. This divergence has been shown within renormalization group to exactly resum in all orders of perturbation theory, producing a non-perturbative restoration of linearity to the spectrum over a wide energy range \cite{KotovInteractions}. 

\begin{figure}[t]
        \centering
        \includegraphics[width=0.8\columnwidth]{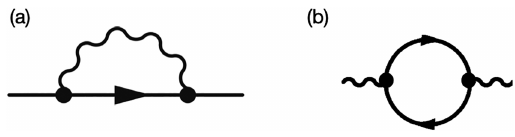}
        \caption{(a)  First order self-energy correction. 
(b) One-loop polarization bubble. Wavy lines represent undressed Coulomb
interactions. Solid straight lines  represent fermion propagators.  
}
         \label{diagrams1}
    \end{figure}

This effect can be understood resorting to standard perturbative renormalization group (RG), as done in QED. The Coulomb interaction Hamiltonian has the form 
\begin{equation}
\mathcal{H}_{\text{C}} = \frac{1}{2}\int \mathrm{d}\mathbf{r}\mathrm{d}\mathbf{r'} n(\mathbf{r})\frac{e^2}{|\mathbf{r}-\mathbf{r'}|}n(\mathbf{r'}),
\label{hamc}
\end{equation}
with $n(\mathbf{r})$ the real space density operator. The free Hamiltonian of semi-Dirac fermions in momentum representation  $\hat{\mathcal{H}}_0(\mathbf{k})$ is given by Eq. (\ref{H_Delta}) at the Lifshitz transition ($\Delta=0$). This Hamiltonian corresponds to the retarded fermionic Green function $\hat{G}_0(\omega,\mathbf{k}) = [\omega -\hat{\mathcal{H}}_0(\mathbf{k})+i\,\mathrm{sgn}(\omega) 0^+]^{-1} $. The self-energy correction of the fermions to leading order in perturbation theory [Fig. \ref{diagrams1}(a)] is \cite{ChubukovInteractions, KotovInteractions}
\begin{align}
\hat{\Sigma}(\mathbf{p})= & \, i\int_{-\infty}^{\infty}\frac{\mathrm{d}\nu}{2\pi}\int\frac{\mathrm{d}^{2}k}{(2\pi)^{2}}\hat{G}_0(\mathbf{k},\nu)V(\mathbf{k}-\mathbf{p})\nonumber \\ 
                                       =  & \, \Sigma_x(\mathbf{p}) \frac{p_x^2}{2m}\sigma_x + \Sigma_y(\mathbf{p}) v p_y\sigma_y,
                                       \label{SigmaMatrix}
\end{align}
where $V(\mathbf{q}) = 2\pi e^2/|\mathbf{q}|$ is the Fourier transform of the Coulomb interaction, and
\begin{align}
\Sigma_x(\mathbf{p}) &=  \frac{\alpha}{4\pi} \log^2 \! \left(\frac{\Lambda}{\varepsilon (\mathbf{p})} \right), \\
\Sigma_y(\mathbf{p}) & =   \frac{\alpha}{\pi} \log \! \left(\frac{\Lambda}{\varepsilon (\mathbf{p})} \right).
\end{align}
Perturbation theory is organized in powers of the parameter $\bar{\alpha} \equiv \alpha/\pi<1$, where $\alpha = e^2/v$ is the dimensionless fine structure constant of graphene \cite{kotov2012} and $\varepsilon(\mathbf{p})\geq 0$ is the positive energy branch of the semi-Dirac fermion dispersion (\ref{SemiDirac_spectrum}).

As anticipated, the effective  inverse mass  has a $\log^2$  divergence,
\begin{equation}
m^{-1}(\omega)  =  m^{-1} \left[1+\frac{\alpha}{4\pi}\log^{2}(\Lambda/\omega)\right].
\label{perturbmass}
\end{equation}
Combining this with the result for the velocity in one loop, namely $v(\omega)=v [1+\frac{\alpha}{\pi}\ln(\Lambda/\omega) ]$, one obtains the RG equations  \cite{KotovInteractions}
\begin{align}
\frac{\mathrm{d}v(\ell)}{\mathrm{d}\ell} & =\,\frac{1}{\pi}v(\ell)\alpha(\ell)= \frac{e^{2}}{\pi}, \label{v1}\\  
 \frac{\mathrm{d}m^{-1}(\ell)}{\mathrm{d}\ell} &=\,m^{-1}(\ell)\frac{\alpha(\ell)}{2\pi}\ell\, , 
\label{vg-rg}
\end{align}
where $\ell\equiv\ln{(\Lambda/\omega)}$ is the RG scale. 
While the velocity is only weakly, logarithmically renormalized (as in graphene) \cite{Gonzalez1994, kotov2012}, the mass is renormalized very strongly, reflecting resummation of leading logarithmic divergences. The solution of the RG equations leads to
\begin{equation}
m^{-1}(\omega)/m^{-1}=\frac{\sqrt{\Lambda/\omega}}{\left[1+\frac{\alpha}{\pi}\ln{(\Lambda/\omega)}\right]^{(\pi/2\alpha)}}.
\label{g-rg-solution}
\end{equation}
Taking the above result on-shell, $\omega=\varepsilon({\bf p})$, one finds for low momenta  $|p_{x}|/\sqrt{2m\Lambda}\ll1$, provided
also $(\alpha/\pi)\ln{(2m\Lambda/p_{x}^{2})}\gg1$, that the quadratic part of the Hamiltonian is renormalized as \cite{KotovInteractions}
\begin{equation}
\frac{p_x^2}{2m} \  \rightarrow \  \frac{p_{x}^{2}}{2 m(\varepsilon)}=\sqrt{\frac{\Lambda}{2m}}\frac{|p_{x}|}{\left[\frac{\alpha}{\pi}\ln{(2m\Lambda/p_{x}^{2})}\right]^{(\pi/2\alpha)}}.
\label{linearization}
\end{equation}
This implies restoration of the spectrum linearity, as in graphene, through the exact resumation of leading logarithmic divergences appearing in higher orders in perturbation theory. The linear spectrum has additional logarithmic modulation visible at the smallest $\alpha$.

\begin{figure*}
            \centering
              \includegraphics[width=0.98\textwidth]{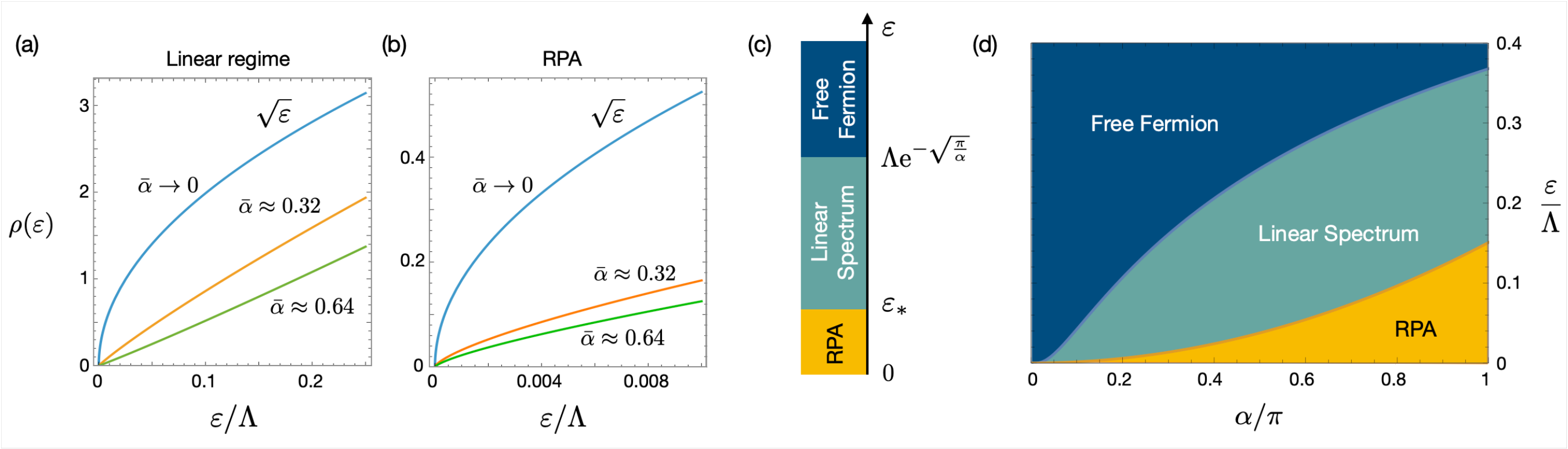}
\caption{(a)-(b): renormalized density of states in units of  $2\pi^2 v/\sqrt{2m}$ versus energy normalized by the ultraviolet cut-off $\Lambda$. (a) linear regime, where there is restoration of the linearity of the spectrum in the non-relativistic direction, and (b) in the ultra-low energy regime (RPA), where screening effects  dominate. Blue curve: bare density of states $\rho\propto\sqrt{\varepsilon}$. Orange: $\bar{\alpha}\equiv \alpha/\pi =0.32$; Green: $\bar{\alpha}\equiv \alpha/\pi =0.64$. (c) Different  weak-coupling regimes in the perturbative RG flow for semi-Dirac fermions. In the free fermion regime, the spectrum is not renormalized. (d) Plot of the boundaries between
different regimes [see Eq. (\ref{window})] versus the perturbative coupling $\alpha/\pi$ for  $N = 4$.  The energy window of the regime with linear spectrum is suppressed in the noninteracting
limit $\alpha\to0$  but extends over a wide energy range for finite coupling $\alpha/\pi<1$ .  The lower bound of the window scales with the number of fermionic species as $N^2$. The RPA-dominated regime takes over the intermediate linear spectrum regime in the large-$N$ limit but remains subdominant to the linear spectrum regime when $N$ is of order 1. 
Adapted from \onlinecite{Elsayed2025}.
  }
\label{fig:PerturbativeRG} 
\end{figure*}

For finite $\alpha<\pi$, the renormalized density of states becomes linear in this regime, $\rho \propto \varepsilon$, over a significant energy range, as shown in Fig. \ref{fig:PerturbativeRG}a.  
This effect   was also found to be present in a larger family of anisotropic Hamiltonians in 2D with generalized semi-Dirac fermions \cite{Elsayed2025}. This indicates that this non-trivial resummation  is robust and not unique to the semi-Dirac case. One also notes that the $2\pi$ Berry phase of the original semi-Dirac particles is a topological property protected from perturbations that do not open a mass gap, and remains unchanged by the renormalization process.

\subsubsection{Weak coupling RPA dominated regime}

As the RG flows, polarization effects eventually recover screening in the infrared at the random phase approximation (RPA) level, reestablishing the original semi-Dirac quasiparticles in the immediate vicinity of the perturbative weak coupling fixed point \cite{ChubukovInteractions, Elsayed2025}. The charge polarization function of semi-Dirac fermions is represented by the bubble diagram in Fig. \ref{diagrams1}b,
\begin{equation}
\Pi(\omega, {\bf p})= -iN \mathrm{tr} \int  \frac{\mathrm{d}^2k}{(2\pi)^2} \frac{\mathrm{d} \nu}{2\pi} \hat{G}({\bf k},\nu)
\hat{G}({\bf k}+{\bf p},\nu+\omega),
\label{polgeneral}
\end{equation}
where $N$ is the fermionic degeneracy. This diagram has been evaluated in approximate analytical form as \cite{wang2017excitonic, ChubukovInteractions}
\begin{eqnarray}
\Pi(\omega,{\bf p}) &&\approx -\frac{N}{2\pi} \frac{\sqrt{2m}}{v} \left[\frac{d_{x} \frac{p_x^2}{2m}}{\left(c_0\frac{p_x^4}{4m^2}+v^2p_y^2 -\omega^2\right)^{1/4}}\right. \nonumber \\
&& \qquad \qquad \qquad \left.+\frac{d_{y}v^2p_y^2}{\left(c_0\frac{p_x^4}{4m^2}+v^2p_y^2 -\omega^2\right)^{3/4}} \right], \qquad
\label{polsd1}
\end{eqnarray}
where $d_x=\sqrt{\pi}\Gamma\left(\frac{3}{4}\right)/4 \Gamma\left(\frac{9}{4}\right)  $, 
$d_y= \sqrt{\pi}\Gamma\left(\frac{5}{4}\right)/4 \Gamma\left(\frac{7}{4}\right) $ and 
$c_0=(16/\pi^2)\left[\Gamma \left(\frac{3}{4}\right)/\Gamma \left(\frac{9}{4}\right) \right]^4$.

The renormalized interaction at the RPA level is $V_{\mathrm{RPA}}(\omega,{\bf p})=V({\bf p})/[1-\Pi(\omega,{\bf p})V({\bf p})]$. 
Within the weak-coupling RPA regime $N\bar{\alpha} \ll 1$ the dominant contribution to screening comes from the static polarization bubble, which has linear scaling with momentum along the non-relativistic direction, $\Pi(0,p_x,0) = -(N/16v) p_x $,  and sublinear scaling  along the relativistic direction, $\Pi(0,0,p_y) = -(N/2\pi) d_y \sqrt{2m/v} \sqrt{p_y} $. The latter contribution is dominant in the $\mathbf{p}\to 0$ limit and sets the characteristic energy $\varepsilon_* = 2m v^2 (N d_y \alpha)^2$ at which $-\Pi(0,0,p_y) V(0,p_y)\sim 1$ and thus the RG solution (\ref{linearization}) crosses over to the screening dominated regime. 

One concludes that Coulomb interactions renormalize the energy spectrum  of semi-Dirac fermions to linear dispersion inside a finite energy window \cite{Elsayed2025}
\begin{equation}
\varepsilon_* < \varepsilon < \Lambda e^{-\sqrt{\pi/\alpha}}.
\label{window}
\end{equation}
The upper bound of the window can be equivalently written as
 $(\alpha/\pi) \ln^2{(\Lambda/\varepsilon)} \sim 1$. For higher energies,  $\varepsilon \gtrsim \Lambda e^{-\sqrt{\pi/\alpha}}$, the log squared contribution to the renormalization of the effective mass (\ref{perturbmass}) is very  weak, and thus the theory is effectively non-interacting. The RG flow for $\alpha/\pi<1$ and $N$ of order 1 can be therefore separated into three different regimes: {\it i)} high energy free fermions; {\it ii)} linear spectrum, where log$^2$  terms resum; and {\it iii)} low energy RPA for $\varepsilon \lesssim \varepsilon_*$. The hierarchy of energy scales in the weak coupling regime is illustrated in Fig. \ref{fig:PerturbativeRG}c. The phase space of each different regime in terms  of energy $\varepsilon$ and  coupling $\alpha$ are shown in panel \ref{fig:PerturbativeRG}d.

Substitution of the RPA, statically screened interaction into Eq. (\ref{SigmaMatrix}) leads to the following self-energy correction to the velocity and effective mass for $\omega \ll \varepsilon_*$,
\begin{align}
v(\omega) &= v \left[1 +\frac{\alpha}{2\pi} \log(\Lambda/\omega)\right], \\
m^{-1}(\omega)&= m^{-1}\left[1-\frac{\alpha}{\pi}\ln\left( d_y N\alpha \right) \ln\left( \Lambda/\omega\right)\right] ,
\label{grpa}
\end{align}
This result is to be compared with the stronger renormalization of the effective mass in Eq. (\ref{perturbmass}). The corresponding RG equations are 
\begin{align}
\frac{\mathrm{d}v(\ell)}{\mathrm{d}\ell} & =\,\frac{1}{2\pi}v(\ell)\alpha(\ell)= \frac{e^{2}}{2\pi}, \\  
 \frac{\mathrm{d}m^{-1}(\ell)}{\mathrm{d}\ell} &=\,-m^{-1}(\ell) \alpha(\ell) \log\left[ d_y \alpha (\ell) N \right]\, ,
\label{vg-rg2}
\end{align}
with $\ell \equiv \log(\Lambda/\omega)$. The velocity is logarithmically renormalized to higher values, and thus the effective coupling $\alpha$ is marginally irrelevant, as in graphene. 
The solution of the RG equation for the effective mass is \cite{ChubukovInteractions, Elsayed2025}
\begin{equation}
m^{-1}(\varepsilon) = m^{-1} \left[ 1 + \frac{\alpha}{2\pi} \log (\Lambda/\varepsilon) \right]^{\gamma(\varepsilon)}
\label{m2}
\end{equation} 
where $\gamma(\varepsilon) = \log\left [ \left( 1+ (\alpha/2\pi) \log (\Lambda/\varepsilon)\right)/(d_y \alpha N)^2\right]$. 

Even though strong non-perturbative many-body effects make semi-Dirac fermions effectively disperse as fully relativistic particles at intermediate energy scales [see Eq. (\ref{linearization}) and (\ref{window})], the  anisotropic nature of these quasiparticles is nevertheless preserved by the renormalization process in the ultra-low energy regime, where the density of states scales as $\rho \propto \sqrt{\varepsilon}$, with strong additional logarithmic corrections to scaling [Eq. (\ref{m2})]. The effect of the spectrum renormalization on the density of states within the weak coupling RPA dominated regime is shown in Fig. \ref{fig:PerturbativeRG}b.

\subsubsection{Type-II semi-Dirac fermions}

Semi-Dirac fermions of type II described by Hamiltonian (\ref{typeII_H}) can have a very different structure of perturbation theory. Coulomb interactions produce a single logarithmic correction to the masses $m$, $m_2$ and to the velocity $v$ both at the Hartree-Fock and at the RPA levels, while generating new terms  in the Hamiltonian \cite{elsayed2026interacting}.  

Retracing the same RG steps in standard perturbation theory, the effective Hamiltonian takes the final form:
\begin{equation}
\hat{\mathcal{H}}(\mathbf{k})=\left ( \frac{k_{x}^{2}}{2m(\varepsilon)} - v(\varepsilon)  k_y \! + \! \alpha \delta \right )\!\sigma_{x} +\left (   \frac{k_{x}k_{y}}{m_2(\varepsilon)} + \alpha c k_{x}  \right )\! \sigma_{y},
\label{ham_eff}
\end{equation}
where 
\begin{equation}
\frac{1}{m(\varepsilon)} =  \frac{L(\varepsilon)}{m},\quad \frac{1}{m_2(\varepsilon)} = \frac{L(\varepsilon)}{m_2},\quad v(\varepsilon) = v L(\varepsilon),
\end{equation}
with $L(\omega)=1+ \frac{\alpha}{\pi}\ln(\Lambda/\omega) $ and $\varepsilon>0$ the bare energy dispersion (\ref{EII}). The constants $\delta \equiv \delta_0(\bar{m}) 2mv^2$ and $c\equiv c_0(\bar{m}) v$ are functions of the ratio of the masses, $\bar{m} \equiv 2m/m_2$, with $\delta_0(1) = 0.150$ and $c_0(1)=0.136$. 

By appropriately relabeling the momentum $k_y 
$, the effective Hamiltonian (\ref{ham_eff}) takes the form of Eq. (\ref{typeII_H}) with a renormalized tuning parameter $\Delta = \alpha (\delta + c  m_2 v)>0$. The Coulomb interaction induces a Lifshitz transition where the semi-Dirac cone renormalizes at low energy to a single anisotropic Dirac cone 
with effective velocity 
$
v_x = \alpha \left[ c+\delta/(m_2 v)\right]
$
along the $k_x$ direction. As in the conventional Lifshitz transition, the energy spectrum crosses over to type-II energy spectrum as the energy increases away from the band crossing.

\begin{figure}[t]
        \centering
        \includegraphics[width=1\columnwidth]{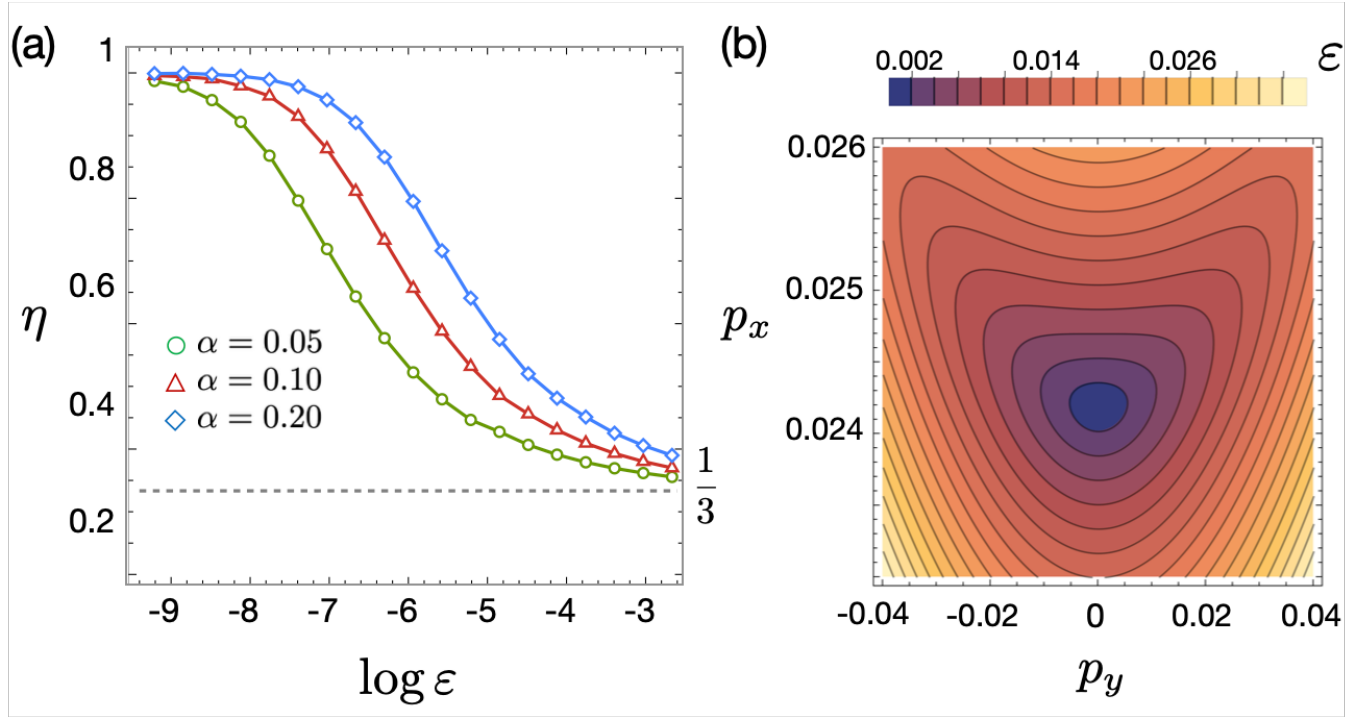}
        \caption{ Scaling of density of states $\rho \propto \varepsilon^{\eta(\varepsilon)}$ and evolution of the Fermi surface of type II semi-Dirac fermions with energy. (a) Scaling exponent  
$\eta(\varepsilon)$ versus $\log \varepsilon$ under the RPA screened Coulomb interaction (connected dots) for $\bar{m}\equiv 2m/m_2=1$. Green, red and blue symbols correspond to  $\alpha=0.05,0.1,0.2$ respectively. (b) Energy evolution of the Fermi surface of the renormalized semi-Dirac cone at Hartree-Fock level for $\bar{m} = 0.5$ and $\alpha=0.2$, following the interaction induced Lifshitz transition.  Momentum in units of $2mv$ and energy in units of $2mv^2$. Adapted from  \cite{elsayed2026interacting}.
}
         \label{Fig:TypeII}
    \end{figure}

This behavior persists in the low energy limit at RPA level, where screening effects dominate. The static polarization function has the exact analytical form \cite{elsayed2026interacting},
\begin{equation}
\Pi(0,\mathbf{p}) = -mN \,\mathrm{max}\!\left[ c_x(\bar{m}) \sqrt{\frac{|p_x|}{2mv}},c_y(\bar{m})\left(\frac{|p_y|}{2mv} \right)^{1/3}\right],
\label{polsd2}
\end{equation}
where $c_x(1)= 0.098, \,c_y(1)= 0.077$. Even though the polarization changes substantially the logarithmic structure of the theory, the overall behavior found at first order remains unchanged. The scaling of the density of states  and the evolution of the shape of the Fermi surface with energy are shown in Fig. \ref{Fig:TypeII}. The scaling exponent  $\eta$ changes from $1$ (characteristic of anisotropic Dirac cone) at low energy towards $\frac{1}{3}$ (free type-II regime) at higher energy.

\subsection{Large $N$ analysis}
\label{sec:largeN}


In the strong coupling regime ($N \alpha \gg 1$) the RG treatment of the semi-Dirac fermion Hamiltonian (\ref{H_Delta}) at $\Delta=0$, controlled by a large $N$ expansion, was found to  exhibit non-Fermi liquid behavior \cite{ChubukovInteractions}. 
In this regime, the effective long-range interaction has the dynamical form $V(\omega,\mathbf{k})\approx-1/\Pi(\omega,{\bf k})$. 
The dynamical self-energy can be written as 
\begin{equation}
\hat{\Sigma}(\omega,\mathbf{k})=\omega \Sigma_0(\omega)\sigma_0 + \Sigma_x(\mathbf{p}) \frac{k_x^2}{2m}\sigma_x  + \Sigma_y(\mathbf{p}) v k_y \sigma_y,
\end{equation} 
where $\sigma_0$ is the identity and $\Sigma_0(\omega) = \partial \hat{\Sigma}(\omega)/ \partial \omega$ determines the quasiparticle residue $Z(\omega)$ via the standard relation $Z^{-1}(\omega)= 1 - \Sigma_0(\omega)$. 
The self-energy has on-shell logarithmic divergences at low energy in leading order: 
$\Sigma_x(\varepsilon) = \gamma_m \ln(\Lambda/\varepsilon)+\Sigma_0$, $\Sigma_y(\varepsilon) = \gamma_v \ln(\Lambda/\varepsilon)+\Sigma_0$,
and $\Sigma_0(\varepsilon) = -\gamma_z\ln(\Lambda/\varepsilon)$, with $\gamma_m$, $\gamma_v$ and $\gamma_z$ proportional to $1/N$.

 Taking into account also vertex corrections, one finds  the RG equations for the mass ($m$), velocity ($v$) and the quasiparticle residue $Z$,
\begin{align}
\frac{\mathrm{d}m(\ell)}{\mathrm{d}\ell}&=-\gamma_m m(\ell), \qquad   \frac{\mathrm{d}v(\ell)}{\mathrm{d}\ell}=\gamma_v v(\ell),    \\
\frac{\mathrm{d}Z(\ell)}{\mathrm{d}\ell}&=-\gamma_z(\ell) Z(\ell).
\end{align}
Here the RG scale $\ell = \ln(\Lambda/\varepsilon)$. The anomalous exponents are calculated to be \cite{ChubukovInteractions}
\begin{equation}
\gamma_m = \frac{0.1261}{N} , \    \   \gamma_v = \frac{0.3625}{N} ,  \  \  
 \gamma_z(\ell) =\frac{\sqrt{15}}{\pi^{3/2}} \ \frac{\ln{[N\alpha(\ell)]}}{N},
\end{equation}
where $\alpha(\ell) = e^2/v(\ell)$ is the running fine structure constant. The coefficient $\gamma_z(\ell)$ governs the decrease of the quasiparticle residue. Integrating the equations down to lowest energies $\varepsilon/\Lambda \ll 1$, one finds 

\begin{equation}
\frac{m(\varepsilon)}{m}=\left(\frac{\varepsilon}{\Lambda}\right)^{\gamma_m}, \  \ 
\frac{v(\varepsilon)}{v}=\left(\frac{\Lambda}{\varepsilon}\right)^{\gamma_v}, \  \ 
\frac{Z(\varepsilon)}{Z}=\left(\frac{\varepsilon}{\Lambda}\right)^{\gamma_z(\varepsilon)}.  \
\label{largenrg}
\end{equation}
At intermediate energy scales, the RG flow  defines a non-Fermi liquid regime, with power-law energy dependence of the quasiparticle residue with energy. The RG analysis on the interplay of Coulomb interactions and quenched disorder is suggestive of the possible existence of instabilities in the strong coupling regime \cite{zhao2016interplay}, although the nature of these states is unclear.

As in graphene, as well as in the previously outlined perturbative regime, interactions flow towards a non-interacting fixed point while the quasiparticle velocity increases. In this weak-coupling region the RG equations have to be modified to account for dynamical RPA dominated screening, and marginal Fermi liquid behavior is recovered \cite{ChubukovInteractions}: 
\begin{equation}
Z(\varepsilon) \sim \frac{1}{\left[ \ln(\Lambda/\varepsilon) \right]^{3/2}}, \  \  \ \varepsilon/\Lambda \ll 1, \  \  \   N \alpha \lesssim 1.
\end{equation}
The mass and velocity renormalization in this latter regime are consistent with the ultra-low energy RPA behavior found within perturbation theory, where $N$ is of order 1 \cite{Elsayed2025}.

\subsection{Many-body effects}
\label{sec:manybody}

The density of states determines the scaling of observables such as the electronic compressibility with density and the electronic specific heat with temperature \cite{Banerjee2012,ChubukovInteractions,Cho2016,KotovInteractions}. Quantum capacitance methods are very sensitive probes to measure  the density of states of low energy quasiparticles and have been used to extract the electronic compressibility in 2D systems \cite{Martin2008, Yu2013}.  

The inverse compressibility is defined as $\kappa^{-1} = n^2 (\partial\mu/\partial n)$, with $n$ the charge density and $\mu$ the chemical potential away from the neutrality point.  The charge response is experimentally determined by the inverse of the density of states $\rho^{-1}(\mu)=\partial\mu/\partial n$ \cite{kotov2012}. For non-interacting semi-Dirac fermions, where $n \propto \mu^{3/2}$, 
\begin{equation} 
\frac{\partial\mu}{\partial n} \propto \frac{1}{\sqrt{\mu}} \propto n^{-1/3}
\end{equation}
 and hence $\kappa^{-1} \propto n^{5/3}$.  In the same way, the low temperature  dependence of the specific heat at fixed volume is  $C_V(T) = -T(\partial^2\mathcal{F}/\partial T^2)_V \propto T^{\frac{3}{2}}$, where  $\mathcal{F}(T) = -T\int \mathrm{d}\varepsilon \rho(\varepsilon) \ln[2+2\cosh(\varepsilon/T)]$ is the  electronic free energy of semi-Dirac fermions, which is specified by the density of states $\rho(\varepsilon)$. 

\begin{figure}[t]
        \centering
        \includegraphics[width=1\columnwidth]{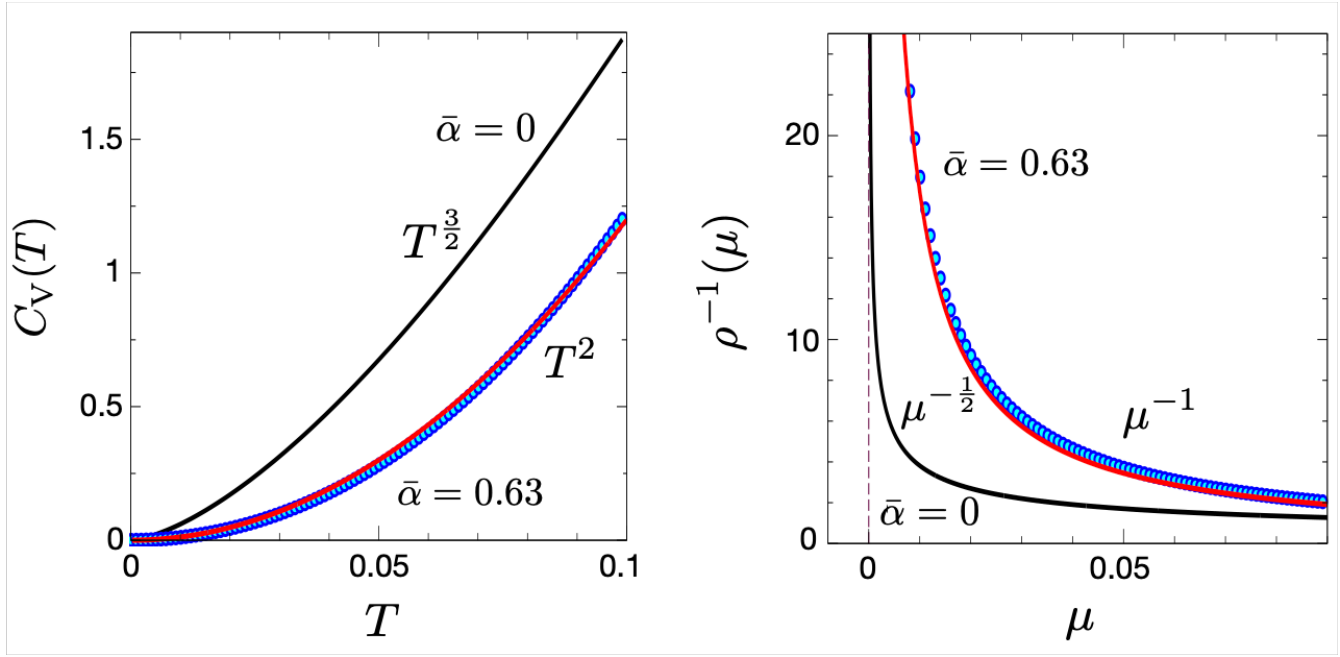}
        \caption{(a)  Specific heat $C_V$ versus temperature  and (b) inverse density of states $\rho^{-1} = \partial \mu/\partial n$ versus chemical potential calculated in the perturbative regime described by Eq. (\ref{v1}) $-$  (\ref{g-rg-solution}) for $\bar{\alpha} = 0.63$ (blue dots). $C_V$  in units of $\sqrt{2m} \Lambda^{3/2}/[(2\pi)^2 v]$. Black curves: non-interacting case ($\bar{\alpha} =0$); red curves:
scaling laws for the purely relativistic case (graphene), $C_V\propto T^2$ and $\partial \mu/\partial n \propto \mu^{-1}$. From \cite{KotovInteractions}}
         \label{Fig:Observables}
    \end{figure}

Following the  discussion in sections \ref{sec:pert} and  \ref{sec:largeN}, Coulomb interactions can produce significant corrections to these and other observables. In the perturbative regime $\bar{\alpha}=\alpha/\pi <1$ with $N$ of order 1, the specific heat scales as $C_V(T) \propto T^2$ (as in graphene) in the energy window (\ref{window}), where there is restoration of the linearity in the energy spectrum. In that regime, $\partial \mu/\partial n \propto \mu^{-1} \propto 1/\sqrt{n}$, and thus the compressibility scales as  $\kappa^{-1} \propto n^{3/2}$. The numerical evaluation of the specific heat and of the inverse density of states in that regime is shown in Fig. \ref{Fig:Observables}. The results are  compared both with the non-interacting case and with the scaling laws of a purely relativistic system.  

In the large $N$ expansion the renormalized density of states scales as \cite{ChubukovInteractions}
\begin{equation}
\rho(\varepsilon) \propto \varepsilon^{1/2 +\phi},
\label{rho3}
\end{equation}
where $\phi = \gamma_v  + \gamma_m/2 = 0.4255/N$.  The leading $1/N$ correction thus increases the density of states exponent away from $\eta = 1/2$, towards linear scaling ($\eta=1$).  Physical observables exhibit modifications of the various power laws.
Specifically, at low temperatures,  the compressibility varies as $\kappa(T) \propto T^{-5/2 - \phi}$, the heat capacity
$C_V(T) \propto T^{3/2 + \phi}$, and the diamagnetic susceptibility $\chi_D(T) \propto - T^{-1/2 - \phi}$. The corresponding scaling behavior for 2D Dirac quasiparticles (non-interacting and with long-range interactions, such as in graphene) can be found in \cite{Sheehy2007,kotov2012}.

Furthermore, Coulomb interactions can produce excitonic instabilities in relativistic systems via chiral symmetry breaking \cite{Pisarski1984, Appelquist1986, Gorbar2002, Khveshchenko2001a, Khveshchenko2004, Khveshchenko2009a, Liu2009a, Gamayun2010}. Semi-Dirac fermions were predicted to undergo a dynamical gap generation at moderately strong Coulomb interactions \cite{wang2017excitonic}, reflecting the  enhancement of the DOS compared to the standard relativistic case (graphene). 

\subsubsection{Quantum hydrodynamics}
\label{sec:hydrodynamics}

The quantum hydrodynamic behavior of semi-Dirac fermions can be unconventional. Quantum hydrodynamics describes the out-of-equilibrium regime where electron relaxation is dominated by interparticle collisions. This occurs when the  collision scattering time $\tau_e$ is much shorter than all other time scales ($\omega \ll \tau_e^{-1}$), including the scattering time due to interactions with phonons or disorder.  For a point-like Fermi surface, where Fermi momentum is zero,  semi-Dirac fermions can have momentum relaxation in the absence of disorder, without violating momentum conservation \cite{Fritz2008}. Transport in the collision dominated regime has universal properties that are controlled by conservation laws \cite{hartnoll2018}.  

The strength of interactions in the hydrodynamic regime   is measured by the ratio between the shear viscosity $\eta$ and the entropy $s$. Lower values indicate stronger interactions and propensity towards quantum viscous flow and possibly quantum turbulence.   It has been conjectured that this ratio has a universal lower bound, $\eta/s\geq \hbar/(4\pi k_B)$, with the equality being satisfied by some strongly coupled field theories \cite{Son2005}.  
Candidate materials to show low viscosity-to-entropy ratios include systems with relativistic particles that lack screening effects \cite{Muller2009}, and thus could be described as strongly interacting quantum fluids.
Possible violations of the bound have been identified in strongly interacting conformal field theories \cite{Brigante2008,Kats2007}, and in holographic gravity models \cite{Hartnoll2016,Baggioli2016}. In the context of quantum solids, violations have been suggested  in nodal materials with strongly anisotropic quasiparticles near the neutrality point, such as semi-Dirac semimetals \cite{Link2018} and nodal-line semimetals \cite{Kim2021a}.

The hydrodynamic transport of semi-Dirac fermions has been calculated in the large $N$ limit, where it was found to be strongly anisotropic. The electric conductivity in the collision dominated regime is insulating in the non-relativistic  direction, $
\sigma_{xx}(T) \propto T^{1/2 +\phi_\sigma}$, and metallic in the other, $\sigma_{yy}(T) \propto T^{-1/2 -\phi_\sigma},
$
with $\phi_\sigma = \gamma_v -\gamma_m/2 = 0.2995/N$ \cite{Link2018}.  

The anisotropy extends to the components of the viscosity tensor, defined by the correlation function $\eta_{\alpha\beta\gamma\delta}(\omega) \propto \mathrm{Im}\langle[\tau_{\alpha\beta} \tau_{\gamma\delta}]\rangle/\omega$, where $\tau_{\alpha\beta}$ is the stress-energy tensor \cite{Bradlyn2012}. It has been found that \cite{Link2018}
\begin{equation}
\eta_{xyxy}(T) \propto T^{5/2 +\phi_\eta},\qquad  \eta_{yxyx}(T) \propto T^{1/2 + \phi_\eta^\prime},
\end{equation}
where $\phi_\eta = (3\gamma_v -\gamma_m/2)/N = 1.0245/N$ and $\phi_\eta^\prime = (3\gamma_m/2 -\gamma_v)/N = -0.1734/N$. From (\ref{rho3}), the entropy scales with temperature as  $s= -(\partial\mathcal{F}/\partial T)_V \propto T^{3/2+\phi}$. Hence, the ratio $\eta_{xyxy}/s \propto T^{1 +\phi_\eta -\phi}$ can be very small at low temperature, in the large $N$ limit, in violation of the conjectured lower bound. 

It remains an open question whether this violation would be present in the perturbative regime, for $N$ of order 1, where the  linearity of the spectrum is restored over a large energy range. In the latter  regime, the viscosity-to-entropy ratio is effectively the one of anisotropic Dirac fermions in graphene, $\eta/s \gtrsim  \hbar/(k_B \alpha^2)$ \cite{Muller2009}.
In any case, semi-Dirac fermions are very good candidates for exhibiting quantum hydrodynamic behavior. Experimental signatures of hydrodynamic transport observed in conventional Dirac fermion systems (graphene) include violation of the Wiedemann-Franz law \cite{Crossno2016}, local negative resistivity \cite{Bandurin2016} and direct observation of viscous flow \cite{Ku2020} and current whirlpools of electrons through imaging \cite{Palm2024}.

\section{Short range interactions and electronic instabilities}
\label{sec:instability}

\subsection{Quantum criticality}
\label{sec:dw}

\subsubsection{Large $N$ expansion}

The quantum critical behavior of relativistic Dirac fermions \cite{Herbut2006, Herbut2009, Assaad2013, Janssen2014}  falls in the universality class of the Gross-Neveu-Yukawa (GNY) theory \cite{Gross1974,Zinn-Justin1991} of chiral symmetry breaking. The most standard framework to describe the QCP of the GNY theory is through a Wilson momentum shell renormalization group approach \cite{Shankar1994}, in which high energy degrees of freedom in the action are integrated out before rescaling all fields, couplings and momenta. 

\begin{figure}[t]
        \centering
        \includegraphics[width=1\columnwidth]{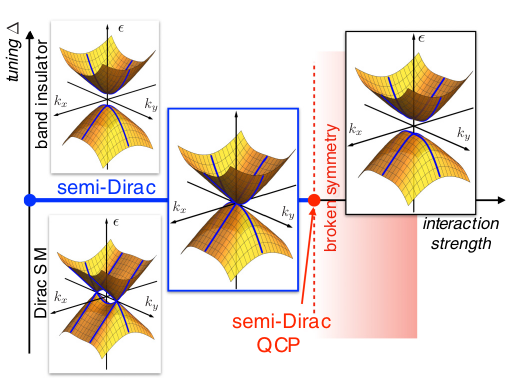}
        \caption{Schematic phase diagram of the band tuning parameter $\Delta$, which sets a topological Lifshitz transition between a Dirac semimetal and a band insulator, versus interaction strength. Sufficiently strong short range interactions produce a QCP separating the semi-Dirac phase $(\Delta=0)$ from a broken symmetry phase.  From \cite{Uryszek2019} }
         \label{Multicritical point}
    \end{figure}

Semi-Dirac fermions symmetry breaking occurs at a multicritical point illustrated in Fig. \ref{Multicritical point}. It consists of a  quantum phase transition governed by interaction strength at a topological Lifshitz transition that is controlled by the tuning parameter $\Delta$. The free action of semi-Dirac fermions ($\Delta=0$) has the form 
\begin{equation}
\mathcal{S}_\psi = \sum_{i=1}^{N} \int_{\vec{k}} \bar{\Psi}_i(\vec{k}) [-ik_0 \sigma_0 + v k_x \sigma_x+ \frac{k_y^2}{2m}  \sigma_y ] \Psi_i(\vec{k}),
\label{Spsi}
\end{equation}
where $\vec{k} \equiv (k_0, k_x, k_y)$, with $k_0$ a Matsubara frequency,  $\int_{\vec{k}} \equiv \int \mathrm{d}\vec{k}/(2\pi)^3$, $\Psi$ is a two-component Grassmann field and $N$ is the number of fermionic flavors.  A four-fermion contact interaction can be decoupled 
into a two-fermion term with a bosonic field 
that represents an order parameter. 
 The simplest instability is a staggered field in the pseudo-spin space, which can be generically called a charge density wave (CDW) order. This instability is described by a scalar field  $\phi(\vec{k})$ that couples to the fermions through the Yukawa coupling 
\begin{equation}
\mathcal{S}_{g} = g \sum_{i=1}^{N} \int_{\vec{k},\vec{q}} \phi(\vec{q}) \bar{\Psi}_i(\vec{k})\sigma_z\Psi_i(\vec{k}+\vec{q}).
\end{equation}
In the ordered phase the semi-Dirac quasiparticles become fully gapped $E(\mathbf{k}) = \pm \sqrt{\varepsilon(\mathbf{k})^2 +|g\phi|^2 }$. In the absence of the fermions, the action of the bosonic field has the general Ginzburg-Landau form  $\mathcal{S}_\phi + \mathcal{S}_\lambda$, where \begin{equation}
\mathcal{S}_\phi = \frac{1}{2}\int_{\vec{q}} (c_0 q_0^2 +c_x q_x^2+c_y q_y^2 +m_\phi^2) |\phi(\vec{q})|^2,  \label{Sphi}
\end{equation}
is the quadratic term, with $m_\phi =0$ at the QCP, and $\mathcal{S}_\lambda = \lambda \int_{\mathbf{r},\tau} |\phi(\mathbf{r},\tau)|^4 $, where $\mathbf{r}$ is the position and $\tau$ the imaginary time. 

Since critical phenomena should not depend on the implementation of cut-offs, it is convenient to treat frequency and momenta on an equal footing in the integrals. Because of the strong anisotropy in the fermionic dispersion, the RG flow is based on integration of high-energy modes from an infinitesimal energy shell  
\begin{equation}
\Lambda \mathrm{e}^{-z \mathrm{d}\ell} < \sqrt{k_0^2 +\varepsilon(\mathbf{k})^2}<\Lambda, \label{Lambda} 
\end{equation}
 where $\Lambda$ is the ultraviolet (UV) cut-off, followed by the rescaling of all frequency and momenta,
\begin{equation}
k_0 = k_0^\prime \mathrm{e}^{-z \mathrm{d}\ell},\quad   k_x = k_x^\prime \mathrm{e}^{-z \mathrm{d}\ell}, \quad  k_y = k_y^\prime \mathrm{e}^{-z_y \mathrm{d}\ell}.
\end{equation}
The nature of the dispersion along the relativistic direction permits rescaling it along with frequency using the same dynamical exponent $z>0$. Invariance of the free  fermionic action (\ref{Spsi})  under the RG requires that the dynamical exponent of the quadratic direction $z_y$ be related to the one in the relativistic direction as $z = 2z_y$ at the tree level. 

Due to the highly anisotropic momentum scaling, one can  choose to keep the  bosonic action coefficient $c_y$  marginal at the tree level, whereas $c_0$ and $c_x$ become irrelevant, with scaling dimension $[c_0]=[c_x]=-z$. 
The fields must then scale with dimensions $[\Psi] = (3z +z_y -\eta_\Psi)/2$ and $[\phi] = (2z + 3z_y  - \eta_\phi)/2$, where $\eta_\Psi$ and $\eta_\phi$ are the anomalous dimensions \cite{Uryszek2019}.   From the above it follows that the couplings $g$ and $\lambda$ have the tree level scaling dimensions $[g] = z/4$ and $[\lambda]=-z/2$. The Yukawa coupling for semi-Dirac fermions is hence a relevant perturbation, whereas the quartic coupling $\lambda$ is irrelevant.

The irrelevance of the coefficients $c_0$ and $c_x$ appears to create difficulties, as the critical bosonic propagator of the theory, $G_\phi(\vec{q}) = 1/(c_y q_y^2)$,  becomes singular  in the $q_y\to 0$ limit inside the UV energy shell (\ref{Lambda}). This unphysical singularity is common in systems with anisotropic scaling  in momentum \cite{Liu2018, Han2018,Han2019} and is fixed through the dressing of the bosonic propagator with the infrared (IR) polarization bubble  in the order parameter channel, $G_\phi^{-1}(\vec{q}) = c_y q_y^2 + \Pi_\mathrm{IR} (\vec{q}) $, where
\begin{equation}
\Pi_\mathrm{IR} (\vec{q}) = \mathrm{tr} \int_{\vec{k}} \sigma_z \hat{G}_\psi(\vec{k}) \sigma_z \hat{G}_\psi(\vec{k} +\vec{q}),\label{Piz}
\end{equation}
and 
\begin{equation}
\hat{G}_\psi(\vec{k}) = \frac{ik_0 \sigma_0 + vk_x \sigma_x + \frac{k_y^2}{2m}\sigma_y }{k_0^2 + v^2k_x^2 +  \frac{k_y^4}{(2m)^2} }
\end{equation} 
is the fermionic propagator. 
This procedure amounts to incorporate Landau damping to the order parameter fluctuations \cite{Chubukov2004}  through gapless particle-hole excitations, and results in a theory that is fully cut-off independent \cite{Uryszek2020}. 
The asymptotic form of the IR bubble for the CDW order is 
\begin{align}\label{IRbubble}
\Pi_{\text{IR}}(\vec{q}) \approx & \, g^2 N \frac{\sqrt{2m}}{v} \left[\frac{a_{x}(q_{0}^{2}+v^{2}q_{x}^{2})}{\left(q_{0}^{2}+v^{2}q_{x}^{2}+b_{y}^{4}\frac{q_{y}^{4}}{4m^2}\right)^{\frac{3}{4}}} \right. \nonumber \\
& \qquad \qquad \left.+\frac{a_{y}\frac{q_{y}^{2}}{2m}}{\left(q_{0}^{2}+v^{2}q_{x}^{2}+b_{y}^{4}\frac{q_{y}^{4}}{4m^2}\right)^{\frac{1}{4}}}\right],
\end{align}
where $a_{x}=\Gamma(5/4)^{2}/\sqrt{2}\pi^{3/2}$, $a_{y}=5\Gamma(3/2)^{2}/16\sqrt{2}\pi^{3/2}$ and $b_{y}=8a_{y}$ \cite{Uryszek2020}.
Since $\Pi_{\text{IR}}(\vec{q})\propto  |q_{y}|$ at  $q_{0},q_{x}=0$,  
it is clear that the self-energy is more relevant in the RG sense than the bare bosonic propagator. Therefore, the dressed propagator becomes $G_\phi(\vec{q}) = \Pi_\mathrm{IR}^{-1} (\vec{q})\propto (g^2N)^{-1}$.  

\begin{figure}[t]
        \centering
        \includegraphics[width=1\columnwidth]{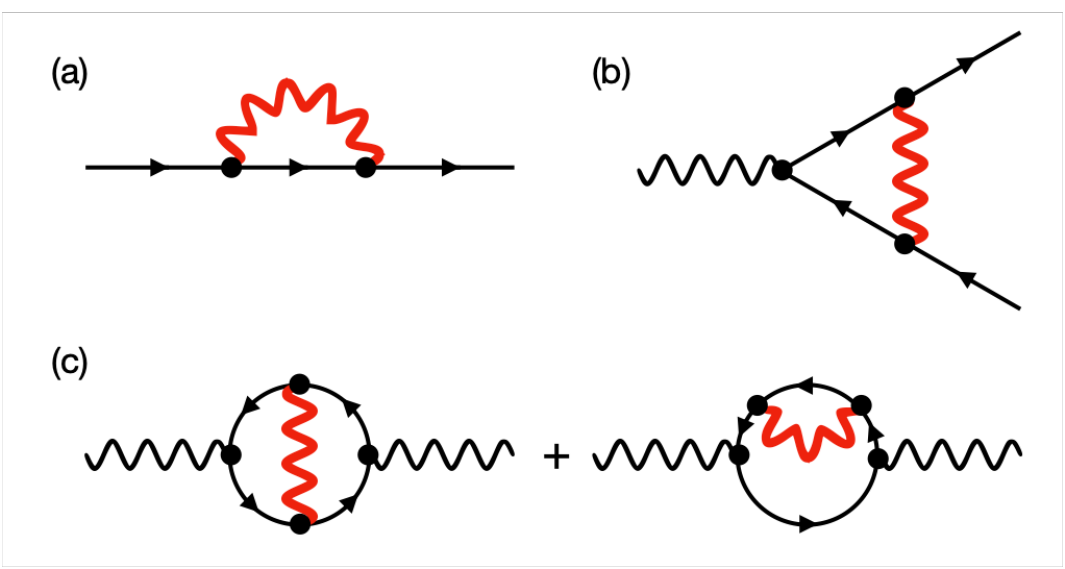}
        \caption{ Diagrams that control the quantum criticality of semi-Dirac fermions near a QCP in large $N$. The red wavy lines represent the bosonic propagator dressed by the IR polarization bubble, which is more relevant than the bare bosonic propagator in the RG sense.  (a) Fermionic self-energy diagram. (b)  Vertex correction  and (c) two-loop corrections to the bosonic propagator, also of order $1/N$. The two-loop diagrams renormalize the bosonic mass and determine the correlation length exponent $\nu_\phi$ at the multicritical point.  Adapted from \cite{Uryszek2020}}
         \label{diagrams}
    \end{figure}

Integration of the high energy modes inside the energy shell (\ref{Lambda}) results in self-energy corrections to the fermions of the form 
\begin{equation}
\hat{\Sigma}(\vec{k}) =[\Sigma_0( k_0 \sigma_0 +vk_x \sigma_x) + ( \Sigma_y k_y^2/2m +\Sigma_\Delta ) \sigma_y] z \mathrm{d}\ell \label{Sigma}\, ,
\end{equation}
represented  in Fig. \ref{diagrams}(a), as well as self-energy corrections $\Delta_\phi z\mathrm{d}\ell$ to the bosonic mass, that are accounted for in the two loop diagrams shown in Fig. \ref{diagrams}(c), both of order $1/N$. The renormalization of the Yukawa coupling $g$ is included through vertex corrections $\Gamma_g z\mathrm{d}\ell$, indicated in Fig. \ref{diagrams}(b). $\Sigma_\Delta$ in Eq. (\ref{Sigma}) renormalizes the tuning parameter $\Delta$ that controls the topological Lifshitz transition.

Absorbing the velocity renormalization as an anomalous dimension $\eta_\Psi= z\Sigma_0$ that enters in the scaling of the fermionic field, the RG leads to a set of equations
\begin{align}
\frac{\text{d}\ln m^{-1}}{\text{d}\ell} & =z-2z_{y}+z(\Sigma_{y}- \Sigma_0) \label{eq:m}\\
\frac{\text{d}\ln g}{\text{d}\ell} & =\frac{z_{y}}{2}+z(\Gamma_{g}-\Sigma_0)-\frac{\eta_{\phi}}{2}\\
\frac{\text{d}\ln m_{\phi}}{\text{d}\ell} & =2z_{y}+z\Delta_{\phi}-\eta_{\phi}=\nu_{\phi}^{-1} \label{nu_phi}\\
\frac{\text{d}\ln\Delta}{\text{d}\ell} & =2z_{y}+z(\Sigma_{\Delta}-\Sigma_0)=\nu_{\Delta}^{-1} \label{nu_Delta}.
\end{align}
$\nu_\phi$ and $\nu_\Delta$ are respectively the correlation length exponents of $m_\phi$ and $\Delta$ at the multicritical point ($m_\phi =0$,  $\Delta =0$). 

Imposing that the mass $m$ does not run, one obtains a relation between the two dynamic exponents in one loop,
\begin{equation}
z-2z_{y}+z(\Sigma_{y}-\Sigma_{0})=0.\label{Z}
\end{equation}
In the same way, as in the usual GNY theory in large $N$, one can scale out the Yukawa coupling through the rescaling $\phi \to \phi/g$ and $m_\phi \to m_\phi g^2$, and hence $g$ should not run. That fixes the bosonic anomalous dimension 
\begin{equation}
\eta_{\phi}=z_{y}+2z(\Gamma_{g}-\Sigma_0), \label{eta_phi2}
\end{equation}
and determines the correlation length exponents
\begin{align}
\nu_{\phi}^{-1} & = z_{y}+z\Delta_{\phi}-2z(\Gamma_{g}-\Sigma_0)\label{nu_phi3}\\
\nu_{\Delta}^{-1} & = 2z_{y}+z(\Sigma_{\Delta}- \Sigma_0)\label{nu_Delta3}.
\end{align}

The exact corrections up to $1/N$ order, numerically calculated using the exact form of the polarization bubble (\ref{Piz}), have been found to be $\Sigma_0=0.0797/N$, $\Sigma_y=0.0214/N$, $\Sigma_\Delta = 0.2755/N$, $\Gamma_g=-0.4350/N$ and $\Delta_\phi = -1.0541/N$ \cite{Uryszek2020}. 
The corresponding critical exponents at the multicritical point are 
\begin{align}
\frac{z_y}{z} = \frac{1}{2} - \frac{0.0292}{N} \, ,
\end{align}
and 
\begin{equation}
\frac{1}{z \nu_{\phi}}  = \frac{1}{2}- \frac{0.0537}{N},\qquad \frac{1}{z \nu_{\Delta}}  = 1+ \frac{0.1958}{N}.
\end{equation}

One can suitably redefine the anomalous dimension $\eta_\phi$ in Eq. (\ref{eta_phi2}) such that $(1/N)^0$ terms are absorbed into the tree level scaling of  $\phi$, while all $1/N$ terms are absorbed into the anomalous dimension $\bar{\eta}_\phi \propto 1/N$. With this choice, $[\phi] = (3z -\bar{\eta}_\phi)/2$, where  $\bar{\eta}_\phi/z = 2\Gamma_g -(\Sigma_0 +\Sigma_y)=  -0.9712/N$. In the same way, we can redefine the scaling of the fermionic fields as $[\Psi] = (7z/2 -\bar{\eta}_\Psi)/2$, such that $\bar{\eta}_\Psi \propto 1/N$, namely,  $\bar{\eta}_\Psi/z = (3\Sigma_0 - \Sigma_y)/2 = 0.1089/N$.  

The universal quantum criticality of the semi-Dirac fermions  GNY theory in large $N$  has been also addressed for spin density wave order and superconductivity \cite{Uryszek2019}, although using only a partial form of the IR bubble (\ref{IRbubble}) to regularize the divergence of the tree level bosonic propagator.

\subsubsection{Other expansions}

Controlled access to the quantum critical behavior of semi-Dirac fermions  via $\epsilon$-expansion, below the upper critical dimension, is subtle and not uniquely defined due to the anisotropic nature of the dispersion. In a generic anisotropic Hamiltonian with point like Fermi surface and $d_L$ linear momentum directions and $d_Q>0$ quadratic momentum directions, short range four fermion interactions become marginal at $2d_L+d_Q=4$ \cite{Sur2019}. Expansion with $d_L=1$ and $d_Q=2-\epsilon$ resulted in non-analytic corrections $\sim \epsilon \ln \epsilon$ to the anomalous dimension near $\epsilon=1$ \cite{Sur2019}, suggesting that semi-Dirac fermions ($d_L=1, d_Q=1$) lie outside the validity of the expansion. 
An alternative expansion with $d_Q=1$ and $d_L= (3 - \epsilon)/2$ \cite{Uryszek2020}  produced a different set of non-analytic corrections to the anomalous dimension. 

Other approaches for local four-fermion interactions include a $1/n$ expansion in the generalized fermionic dispersion $k_y^{n}$ along the non-relativistic direction in 2D, with $n=2$ the semi-Dirac case \cite{RoyFoster}.  This approach describes a controlled ascent from one dimension ($n\to \infty$), where the RG calculation is organized in powers of $\epsilon=1/n$. The density of states of the engineered Hamiltonian vanishes as $\rho (\varepsilon) \sim |\varepsilon|^{1/n}$ at the neutrality point, and hence interactions are irrelevant at weak coupling. In analogy with conventional $\epsilon$-expansions, the small parameter $\epsilon$ enters as the engineering dimension of the interactions, which have scaling dimension $-\epsilon$ and are marginal at $\epsilon=0$. This setup corresponds to an effective dimension $d_*=1+\epsilon$, with additional control of the fluctuations  along the $k_y$ direction through the $1/n$ expansion. The two parameters $\epsilon$  and $1/n$ are treated independently in the RG scheme and are set to be equal at the end.   

Interactions give rise to either a fluctuation driven first order phase transition or else to a broken symmetry state in the strong coupling regime \cite{RoyFoster}. The latter occurs through a continuous transition at a quantum multicritical point, where the anisotropic semimetal, band insulator and ordered phases meet. The dominant instabilities are  CDW, spin density wave order and $s$-wave superconductivity. This approach retains key features of 1D physics, such as spin-charge separation. It is not presently clear to what extent this method faithfully describes the physics of interactions for semi-Dirac fermions ($\epsilon=1/2$) at effective dimension $d_*=3/2$.

\subsection{Superconductivity}
\label{super}

Superconductivity can be generically considered in the case of an arbitrary Hamiltonian $\hat{\mathcal{K}}(\mathbf{p})$  with semi-Dirac quasiparticles that preserves time reversal symmetry, namely $\mathcal{T}\mathcal{\hat{K}}(\mathbf{p})\mathcal{T}^{-1} =\mathcal{\hat{K}}(\mathbf{p})$, where $\mathcal{T}$ is the time reversal symmetry operator and $\mathbf{p}$ the momentum measured from the center of the BZ. The general  Bogoliubov deGennes (BdG) Hamiltonian in the Nambu basis $\{{|\mathbf{p},\uparrow\rangle, |-\mathbf{p},\downarrow\rangle}\}$ 
 that describes pairing between semi-Dirac points sitting in opposite sides of the BZ, such that Cooper pairs have zero total momentum, is
\begin{equation}
\hat{\mathcal{H}}_{\text{BdG}}(\mathbf{p})=\left(\begin{array}{cc}
\hat{\mathcal{K}}(\mathbf{p}) & \hat{\Delta}_s\\
\hat{\Delta}_s^{*} & -\hat{\mathcal{K}}(\mathbf{p})
\end{array}\right)\label{eq:Ham}.
\end{equation}
$\hat{\Delta}_s$ is a 2$\times$2 matrix that describes different pairing channels in the orbital space. Intra-orbital pairing $\hat{\Delta}_s = \Delta_s \sigma_0$ produces a fully gapped state 
in the singlet channel \cite{Uchoa2017}.   

Expanding $\hat{\mathcal{K}}(\mathbf{p})$ in the vicinity of a semi-Dirac point gives the BdG Hamiltonian of the fully gapped state in the continuum limit 
\begin{equation}
\hat{\mathcal{H}}_{\mathrm{BdG}}(\mathbf{k}) =    \frac{k_x^2}{2m} \sigma_x \otimes \tau_z +  vk_y \sigma_y  \otimes \tau_z+ \Delta_s \sigma_0 \otimes \tau_x\,,\label{BdG2}
\end{equation}
where $\tau_i$  $(i=x,y,z)$ are the standard Pauli matrices in the Nambu space and $\mathbf{k}$ is the momentum away from the semi-Dirac point. We assume a suitable gauge such that $\Delta_s$ is real. Expansion of $\hat{\mathcal{K}}(\mathbf{p})$ around the opposite semi-Dirac point gives an equivalent Hamiltonian for the other valley, which is accounted for as a degeneracy. Diagonalization of Eq. (\ref{BdG2}) gives the energy spectrum
$
 E(\mathbf{k}) = \pm \sqrt{\varepsilon^2(\mathbf{k})+\Delta_s^2}
$,
where $\varepsilon(\mathbf{k})$ is the semi-Dirac spectrum (\ref{SemiDirac_spectrum}). 

As in the case of Dirac fermions \cite{AntonioTMD, Khveshchenko2001b}, the suppression of the density of states at the semi-Dirac point implies that superconductivity is quantum critical when the Fermi surface is point-like.
The free energy of the superconducting state has the form 
$
\mathcal{F}= -T\sum_{\mathbf{k},s}\ln[2+2\cosh(E_\mathbf{k}/T)]+ \Delta_s^2/g,
$
 where $g>0$ is the effective local attractive interaction that stabilizes Cooper pairs and $T$ is temperature. 

\begin{figure}[t]
        \centering
        \includegraphics[width=0.85\columnwidth]{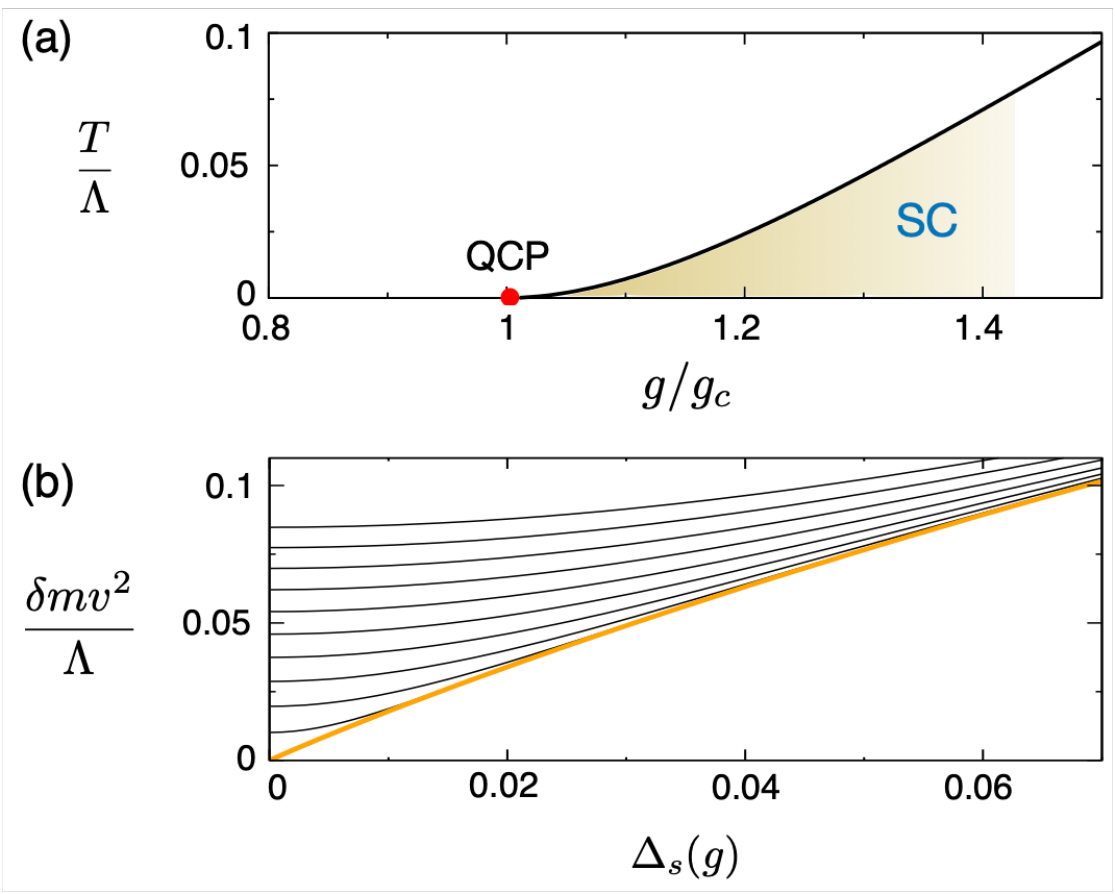}
        \caption{(a) Phase diagram temperature versus coupling strength $g$. Colored region indicates a gapped superconducting phase. The red dot at $g=g_c$ is a QCP. In its vicinity, the gap scales as $\Delta_s \propto (1-g_c/g)^2$ at the mean-field level. (b) Scaling of the Meissner kernel anisotropy $\delta = Q_x/Q_y$  times $mv^2/\Lambda$ versus $\Delta_s(g)$ for fixed temperature. $T/\Lambda$ ranges from zero (orange line) to 0.025 in 0.0025 steps. From \cite{Uchoa2017}}
         \label{Gap}
    \end{figure}

 After minimization of the free energy in $\Delta_s(T,g)$, the gap scales with the coupling in the $T\to0$ limit as
 \begin{equation}
 \Delta_s(0,g) = \Lambda/\gamma_1^2 (1-g_c/g)^2 \theta(g-g_c),
 \end{equation}
where $g_c = 1/(\sqrt{\Lambda} \rho_0)$ is the cut-off dependent critical coupling, $\rho_0$ is the prefactor of the density of states (\ref{DOS-I}), namely $\rho(\varepsilon) = \rho_0 \sqrt{\varepsilon}$, and $\gamma_1 = \Gamma^2\left(\frac{3}{4}\right)/\sqrt{\pi} \approx 0.85$. The mean field critical temperature is $T_c(g) \approx \gamma_1^2 \Delta_s(0,g)$, shown as a phase boundary in Fig. \ref{Gap}(a). In the vicinity of $T_c$ and away from the QCP the gap scales with temperature following the standard BCS form $\Delta_s(T,g) \approx 2.02 \Delta_s(0,g)\sqrt{T_c/T-1}$.  The specific heat jump at $T_c$ normalized by the specific heat in the normal side of the transition has a universal value $\delta C_V \approx 0.71$. This value is intermediate between the case of Dirac fermions in 2D, where  $\delta C_V \approx 0.35$ \cite{Uchoa2005}, and the Fermi liquid case  $\delta C_V \approx 1.43$ \cite{Tinkham2004}.

One can incorporate condensate flow with momentum $\mathbf{q} = (q_{x}, q_{y})$  into the free energy through the momentum substitution $\mathbf{k} \to \mathbf{k}+ \mathbf{q}$. Expanding the free energy in powers of $\mathbf{q}$ and $\Delta_s$ in the zero temperature limit, one obtains the  Ginzburg-Landau free energy for semi-Dirac fermions describing the superconducting QCP.  As typical of quantum phase transitions, the related Ginzburg-Landau free energy per node is not analytic and has the form \cite{Uchoa2017}
\begin{equation}
\mathcal{F}_{\textrm{GL} }= \frac{ \gamma_x q_x^2}{\sqrt{m}v} |\Delta_s|^\frac{3}{2} +\gamma_y \sqrt{m}v q_y^2 \sqrt{|\Delta_s|} + r(g) \Delta_s^2 +u|\Delta_s|^{\frac{5}{2}},
\label{GL}
\end{equation}
 where $\gamma_x$ and $\gamma_y$ are numerical constants, $u>0$ and $r(g) = r_0 \delta g$, with $r_0 <0$ and $\delta g \equiv 1 - g_c/g$ the reduced coupling near the QCP. 
 
 At finite magnetic field the condensate momentum is proportional to the vector potential, $\mathbf{q} = -(2e/\hbar c) \mathbf{A}$, where $c$ is the speed of light. The supercurrent $\mathbf{j} = c\partial\mathcal{F}_\mathrm{GL}/\partial \mathbf{A}$ is strongly anisotropic at $T=0$ near the QCP, $j_i = Q_i A_i $ ($i=x,y$), where
 \begin{equation}
 Q_x \propto - \frac{e^2}{\hbar^2c} \frac{\Delta^\frac{3}{2}}{\sqrt{m}v} ,\qquad  Q_y \propto - \frac{e^2}{\hbar^2c} \sqrt{\Delta}\sqrt{m}v 
 \label{Q}
 \end{equation}
 are the components of the London Kernel. The scaling of the London Kernel anisotropy $\delta = Q_x/Q_y$ with $\Delta_s$ is shown in Fig. \ref{Gap}(b) for different temperatures.  In the vicinity of the critical temperature, $T\approx T_c$, the Ginzburg-Landau free energy has the usual form, with quadratic and quartic terms in $\Delta_s$. The London Kernel of semi-Dirac fermions remains anisotropic near $T_c$ but recovers the standard BCS scaling with the gap,  $Q_i\propto \Delta_s^2$.
 
 In anisotropic superconductors the supercurrent is not generically parallel to the vector potential. Conservation of current is expressed through the transversality condition 
 $\mathbf{k} \cdot \mathbf{j}=0$, based on which Eq. (\ref{Q}) is consistent with the choice of a suitable fixed gauge that sets the longitudinal component of supercurrent to zero. Gauge invariance is recoverable through the incorporation of screening effects via virtual optical plasmons \cite{Pines}. 
In superconducting thin films with finite thickness, optical plasmons can screen in the London limit $\mathbf{q}\to0$ for any small offset of the chemical potential $\mu$ away from the semi-Dirac point.  Fingerprints of the QCP are observable in the strong coupling regime  $0<|\mu| \ll T\ll |\Delta_s| $ through the scaling of physical observables with temperature and the gap.   
  
The strong anisotropy in the dispersion can produce an unconventional Meissner response, where both the penetration depth and the coherence length have distinct scaling exponents along each direction in the vicinity of the QCP.   In semi-Dirac fermion superconductors with uniaxial anisotropy such as strained graphene,  the total London kernel is equal to the contribution of a single nodal point times the nodal degeneracy. The penetration depth for a thin film of thickness $d$ is $\lambda_i = \sqrt{-cd/(4\pi Q_i)}$. Near the QCP $\lambda_x(g) \propto |\delta g|^{-\frac{3}{2}}$ and $\lambda_y(g) \propto |\delta g|^{-\frac{1}{2}}$. In the opposite limit, near the critical temperature $T_c$, the penetration depth follows the conventional BCS scaling with the gap, $\lambda_i \propto \Delta_s^{-1}(T)$. 
  
 The zero temperature coherence lengths $\xi_i$ can be extracted by dimensional analysis of the  free energy (\ref{GL}) \cite{Uryszek2019}, 
 \begin{equation}
 \xi_x^{-2} |\Delta_s|^{\frac{3}{2}}  \simeq \xi_y^{-2}\sqrt{|\Delta_s|} \simeq \delta g \Delta_s^2.
 \end{equation} 
 Thus the coherence lengths diverge near the QCP with different exponents, $\xi_x\propto |\delta g|^{-1}$ and  $\xi_y\propto |\delta g|^{-2}$. The ratio between the penetration depth and  coherence length $\kappa_i =\lambda_i/\xi_i$ indicates that the order parameter is soft for variations in the direction where the quasiparticles have parabolic dispersion, $\kappa_x \propto (\delta g)^{-\frac{1}{2}}\gg 1$, and rigid for variations along the direction of linear dispersion, $\kappa_y \propto (\delta g)^{\frac{3}{2}}\ll 1$. 
The penetration of a magnetic flux in a superconductor of semi-Dirac fermions may stabilize a smectic state with stripes of superconducting domains intercalated by thin normal strips of magnetic flux, rather than vortices \cite{Uchoa2017}. This state is distinct from type-1.5 superconductivity, where competing coherence lengths lead to phase separation into vortex stripes and vortex clusters \cite{Moshchalkov}.

\section{Future directions}
\label{sec:Conclusion}
 
Semi-Dirac fermions are exotic 2D quasiparticles that emerge through topological Lifshitz transitions of Fermi surfaces, points or lines. Their  formation at the merger of two or more Dirac cones is not limited to electrons, but can also occur in a variety of systems and platforms, including cold atom gases, optical waveguides, and in frustrated magnetic lattices with quantum spin liquids. 
The universal nature of this phase transition raises the intriguing possibility that these exotic particles could emerge not only in systems exhibiting topological order,  in the form of semi-Dirac Majorana fermions
, but also in the context of other many-body instabilities merging effective low energy Dirac cones. 

From a different perspective, if one starts with a gas of semi-Dirac fermions, could one experimentally reach the integer and fractional quantum Hall regimes at strong magnetic field? The most direct electronic evidence for semi-Dirac fermions currently comes from the 3D nodal-line semimetals ZrSiS~\cite{BasovZRSIS} and SrAs$_3$~\cite{jeon2025}, where the semi-Dirac dispersion occurs on critical 2D momentum slices through a 3D band structure. While the 2D integer quantum Hall effect may be experimentally realized in 3D systems with nodal lines~\cite{yin2019}, the intrinsically interacting nature of the fractional quantum Hall effect would very likely require precise tuning of the carrier density.  A major experimental goal is therefore to realize semi-Dirac fermions in high-mobility, gate-tunable, 2D electronic systems.

 Light-matter coupling in systems with semi-Dirac fermions remains a wide open area of research. Optical properties of solids are determined by the permittivity tensor, which rules the collective response of electrons (plasmons) to electromagnetic waves \cite{Pines1999}.
It is known that strong 2D quasiparticle anisotropy can  produce directional, tunable, hyperbolic plasmons~\cite{nemilentsau2016,vanveen2019, venturi2024, ruta2025}, which are expected in hyperbolic media, where the components of the permittivity tensor in orthogonal directions have opposite signs. Hyperbolic materials have very large photonic density of states, support electromagnetic waves with very large wavevectors and have negative index of refraction \cite{Poddubny2013,Basov2016}. Hyperbolic materials also support ultra-stong fluctuating vacuum electric fields that were demonstrated to alter properties of proximal materials in layered heterostructures \cite{Keren2026}. These features make systems with semi-Dirac fermions attractive for optoelectronic purposes, as well as for the task pertaining to control of other materials via engineering of vacuum fluctuations. 

Near-field optical imaging in exfoliated ZrSiSe crystals has revealed hyperbolic plasmon propagation in a nodal-line metal~\cite{shao2022a} that has electronic structure similar to ZrSiS~\cite{hu2016}.
In the same venue, the plasmon-polaritonic response of semi-Dirac fermions, where plasmons form a bound state with light, is relatively unexplored. 
Recent experiments in layered black phosphorus have demonstrated the existence of transient, out-of-plane, hyperbolic plasmon polaritons~\cite{fu2024} and provided spectroscopic evidence for pressure-induced in-plane hyperbolic plasmons~\cite{liu2026}.
These experiments are based on systems where the presence of semi-Dirac fermions has not been demonstrated yet. 
The field of optoelectronics for semi-Dirac fermions is ripe for advancements and is still in its infancy.

As Dirac cones with different winding numbers merge, inversion symmetry is broken locally near the semi-Dirac point, permitting formation of a finite Berry curvature dipole, and possibly higher order Berry curvature multipoles. This property suggests that semi-Dirac fermion systems are promising platforms for probing band geometry through non-linear Hall measurements \cite{jiang2025}. Non-linear Hall conductivity and rectified photocurrents have emerged as powerful diagnostics of band topology, Berry curvature, and spatially inhomogeneous quantum responses \cite{ma2021,ma2023}. These methods can thus provide sensitive probes of the Lifshitz tuning parameter, chemical potential, and Berry curvature symmetry breaking in bulk, and spatially resolve the contribution from edge states in topological phases.

The electronic transport of semi-Dirac fermions in bulk, in the metallic phase, reflects the anisotropic nature of these quasiparticles and may have very interesting physics in the presence of strong disorder.  Anderson localization for semi-Dirac fermions could be unusual and remains  to be theoretically explored. Strong localization for Galilean invariant electrons is determined by three main universality classes of disorder based on basic symmetries of the system under time reversal and spin rotations \cite{Evers2008}. The number of universality classes for 2D Dirac fermions  was extended to include additional chiral (pseudospin) and particle-hole symmetries. Semi-Dirac fermions are neither fully Galilean invariant nor fully relativistic particles and their fate regarding localization under different types of disorder is presently unclear.   

While the presence of a Berry curvature dipole reflects single-particle physics, the many-body problem of semi-Dirac fermions is surprisingly rich compared to the relativistic case in 2D.  Coulomb interactions were theoretically predicted to produce strong renormalization effects in both the perturbative and strong coupling regimes.   
 As in graphene, short range interactions can by their turn produce different strong coupling  instabilities within the charge, spin and valley sectors. However, the essential anisotropy in the kinetic energy of semi-Dirac fermions brings technical challenges to controlled many-body expansions near the onset of competing instabilities in the strong coupling regime. The nature of the dominant many-body states is not yet clear and remains an open question. 
Furthermore, semi-Dirac fermions are prone to exhibit quantum hydrodynamic behavior in the collision dominated regime, perhaps even more so than Dirac fermions in graphene, and show anisotropic scaling near quantum critical points that could stabilize emergent modulated phases.  

 All of these raise the question of what promising new directions could result from the interplay of  interactions and strong quasiparticle anisotropy with  quantum geometry \cite{Yu2025}. For instance, it has been realized that many-body bound states and collective modes can have internal quantum structure reminiscent of the band geometry of the original single particle states.  Bound states of light with plasmons and  excitons in electronic bands with non-trivial quantum geometry can have a finite quantum geometric dipole moment \cite{Cao2021a, Xie2024a,Xie2026, Lozano2025}. The implications of the internal quantum structure of polaritonic bound states in hyperbolic media, including in systems with semi-Dirac fermions, have yet to be investigated.

There are plenty of unexplored research avenues in the broad subject of semi-Dirac fermions in quantum matter,  both in experiments and theory. This review has highlighted some prominent features of these exotic quasiparticles, their current experimental status, and leaves a set of open questions to the future.

\section*{Acknowledgments}
We are indebted to A. Chubukov, J. Herzog-Arbeitman, F. Kruger, S. Parmeswaran and O. Sushkov for stimulating discussions.
We are grateful to G. Montambaux 
for insightful comments and suggestions regarding the manuscript.
BU acknowledges NSF grant No. DMR-2529526 for support. Y.S. acknowledges support from the U.S. Department of Energy, Office of Science, Basic Energy Sciences, under Award No. DE-SC0026357 for studies of light-matter interactions in quantum materials. Work at Columbia is supported under NSF Awards No. DMR-2614607 and DMR-2210186.

\bibliography{Refs_RMP}

\end{document}